\documentclass[preprint2]{aastex}

\newcommand{\mem}[1]{\ensuremath{\mathrm{ #1}}}
\newcommand{\nezw}{\ensuremath{^{22}\mem{Ne}}}
\newcommand{\nadr}{\ensuremath{^{23}\mem{Na}}}
\newcommand{\spr}{\mbox{$s$-process}}
\newcommand{\sprn}{\mbox{$s$ process}}
\newcommand{\mgfu}{\ensuremath{^{25}\mem{Mg}}}
\newcommand{\n}{\ensuremath{\mem{n}}}
\newcommand{\kap}[1]{Sect.\,\ref{#1}}
\newcommand{\msun}{\ensuremath{\, {\rm M}_\odot}}
\newcommand{\abb}[1]{Fig.\,\ref{#1}}
\newcommand{\jahre}{\ensuremath{\, \mathrm{yr}}}
\newcommand{\lsun}{\ensuremath{\, {\rm L}_\odot}}
\newcommand{\tab}[1]{Table\,\ref{#1}}
\newcommand{\kelv}{\ensuremath{\,\mathrm K}}
\newcommand{\komma}{\mathrm{ \hspace*{0.1cm},}}
\newcommand{\xeac}{\ensuremath{^{128}\mem{Xe}}}

\slugcomment{DRAFT: \today, LA-UR-05-8084}

\shorttitle{He-shell flash convection}
\shortauthors{Herwig \etal.}

\begin{document}

\title{Hydrodynamic simulations of He-shell flash convection} \author{
Falk Herwig\altaffilmark{1}\altaffiltext{1}{Los Alamos National
Laboratory, Los Alamos, NM 87544, e-mail: fherwig/rmhx/timmes@lanl.gov}, 
Bernd Freytag$^1$ \altaffilmark{2}\altaffiltext{2}{Department of Astronomy
and Space Physics at Uppsala University, Sweden, e-mail: bf@astro.uu.se} 
\altaffilmark{3}\altaffiltext{3}{Department of Physics and Astronomy,
Michigan State University, East Lansing, MI 48824}, 
Robert M. Hueckstaedt$^1$, Francis X. Timmes$^1$}

\begin{abstract}
  We present the first hydrodynamic, multi-dimensional simulations of
  He-shell flash convection. Specifically, we investigate the
  properties of shell convection at a time immediately before the
  He-luminosity peak during the $15^\mem{th}$ thermal pulse of a
  stellar evolution track with initially two solar masses and
  metallicity $Z=0.01$. This choice is a representative example of a
  low-mass asymptotic giant branch thermal pulse. We construct the
  initial vertical stratification with a set of polytropes to resemble
  the stellar evolution structure. Convection is driven by a constant
  volume heating in a thin layer at the bottom of the unstable layer.
  We calculate a grid of 2D simulations with different resolutions and
  heating rates. Our set of simulations includes one low-resolution 3D
  run.  The computational domain includes $11.4$ pressure scale
  heights. He-shell flash convection is dominated by large convective
  cells that are centered in the lower half of the convection
  zone. Convective rolls have an almost circular appearance because
  focusing mechanisms exist in the form of the density stratification
  for downdrafts and the heating of localized eddies that generate
  upflows. Nevertheless, downdrafts appear to be somewhat more
  focused. The He-shell flash convection generates a rich spectrum of
  gravity waves in both stable layers above and beneath the convective
  shell. The magnitude of the convective velocities from our 1D
  mixing-length theory model and the rms-averaged vertical velocities
  from the hydrodynamic model are consistent within a factor of a
  few. However, the velocity profile in the hydrodynamic simulation is
  more asymmetric, and decays exponentially inside the convection
  zone. An analysis of the oscillation modes shows that both g-modes
  and convective motions cross the formal convective boundaries, which
  leads to mixing across the boundaries.  Our resolution study shows
  consistent flow structures among the higher resolution runs, and we
  see indications for convergence of the vertical velocity profile
  inside the convection zone for the highest resolution simulations. Many of
  the convective properties, in particular the exponential decay of
  the velocities, depend only weakly on the heating rate. However, the
  amplitudes of the gravity waves increase with both the heating rate
  and the resolution. 
\end{abstract}


\section{Introduction}
\label{kap:intro}

We present hydrodynamic simulations of the interior shell
convection zone driven by the He-shell flash in thermal pulse
Asymptotic Giant Branch (AGB) stars. 

\subsection{AGB evolution}
AGB stars are the final evolution stage of low- and intermediate mass
stars before formation of a white dwarf \citep{iben:83b}.  Nuclear
production in AGB stars contribute to the chemical evolution of
galaxies \citep[for example][]{travaglio:04,renda:04}. The chemical
yields are support the interpretation stellar abundance from
dwarf-spheroidal galaxy satellites of the Milky Way relative to
stellar abundances of galactic halo stars \citep{venn:04,geisler:05}.
This may provide important clues about the cosmological
origin of our galaxy. Another role of AGB
stars have recently been discussed as a potential source of the
abundance anomalies observed in globular cluster star members
\citep{ventura:02,denissenkov:03a}, a question that is still debated.

AGB stars produce substantial amounts of Li, C,
N, \nezw, \nadr, and the neutron-heavy Mg isotopes.  These stars are
the progenitors of planetary nebulae nuclei and white dwarfs,
including those thought to orbit C-rich extremely metal-poor stars
with \spr\ signature \citep[CEMP-s, ][]{beers:05}. The favoured
interpretation of their strongly non-solar abundance pattern is the
pollution of the EMP star with material transfered from the white dwarf
progenitor AGB star. This interpretation is largely based on the fact
that the abundance patterns of CEMP-s stars in general resemble the C- and
\spr\ overabundance predicted for EMP AGB stars. However, very
significant discrepancies between observed and predicted abundances
exist which could indicate some serious problem with current 1D
stellar evolution and nucleosynthesis simulations. Most likely those
problems are related to the 1D approximate treatment of mixing
that originate from multi-dimensional, convective fluid motions.

Especially at extremely low metallicity, AGB stars can produce many
species in a primary mode, i.e.\ only with the H and He available from
the Big Bang \citep[for example]{siess:02,herwig:04a}.  AGB stars are
also the nuclear production site of the main component of the \sprn\
\citep{busso:99}, which produces roughly half of all trans-iron
elements \citep{arlandini:99}. The interpretation of high-precision
laboratory measurements of isotopic ratios of heavy elements in
pre-solar meteoritic SiC grains depends sensitively on the
thermodynamic and the closely related mixing conditions at the bottom
of the He-shell flash convection zone
\citep{zinner:98,lugaro:02b,lugaro:02a}. In this particular layer the
$\nezw(\alpha,\n)\mgfu$ reaction releases neutrons very rapidly, and
activates many s-process branchings. Experimental measurements of
nuclear properties of the unstable branch-point nuclei are now
underway \citep{reifarth:05} and in conjunction with grain
measurements will provide new constraints for mixing at convective
boundaries in the future.

Currently, only 1D stellar evolution models exist to describe the
processes of the AGB stellar interior quantitatively. However, while
spherical symmetry is globally a very good approximation, a detailed
description of mixing processes induced by the hydrodynamic
fluid-flows requires multi-dimensional simulations. Only over long
times, equivalent to many convective turn-overs, can they be described
accurately by the average quantities provided by the commonly used
local theories, like the mixing-length theory
\citep{boehm-vitense:58}. However, it is well known that these
theories do not describe the quantitative properties of mixing
at convective boundaries. For these reasons, we start our
investigations of the hydrodynamic properties of stellar interior
convection with the He-shell flash convection in thermal pulse AGB
stars.

\subsection{Hydrodynamic simulations of stellar convection and application to 1D stellar evolution calculations}
Most multi-dimensional simulations of stellar convection address the
outer convection zones. For example the solar convection zone
including the surface granulation \citep{nordlund:82,stein:98},
convection below the surface \citep{stein:89} or the role of magnetic
fields and rotation in the solar convection zone \citep{brun:04b}
Other simulations include work on deep envelope convection in giants
\citep{porter:94,porter:00,freytag:02,robinson:04}, shallow surface convection, for example in A stars and white dwarfs
\citep{freytag:96} and the general properties of
slab convection \citep{hurlburt:86}. Hydrodynamic simulations of
stellar interiors include an investigation of semi-convection
in massive stars by \citet{merryfield:95} and core convection in rotating
A-type stars \citep{browning:04}.

The work by \citet{freytag:96} motivated \citet{herwig:97} to
include the exponential overshooting derived from hydrodynamical
simulations to all convective boundaries. The e-folding distance for
convective boundaries in the stellar interior was derived
semi-empirically for main-sequence stars, and then applied to AGB
stars \citep{herwig:99a}. Similar approaches were considered by
\citet{mazzitelli:99}, \citet{ventura:00}, \citet{cristallo:04}, and
\citet{althaus:05}.

However, convection in shells driven by nuclear
energy release is different than envelope convection. Typically, the
Mach numbers are smaller throughout the convection zone, and the
bordering radiative layers are relatively more stable. Accordingly, the
semi-empirical determination of the overshooting value
$f_\mem{ov}$\footnote{Here, the overshoot formulation is $ D_{\rm OV}
= D_0 \exp{\left( \frac{-2 z}{f_\mem{ov} \cdot H_{\rm p}}\right)}$,
where $D_0$ is the mixing-length theory mixing coefficient at the base
of the convection zone, $z$ is the geometric distance to the
convective boundary, $H_{\rm p}$ is the pressure scale height at the
convective boundary, and $f_\mem{ov}$ is the overshooting parameter
\citep{herwig:97}.} gives smaller values for the stellar interior ($f_\mem{OV} < 0.016$)
compared to the shallow surface convection ($0.25 < f_\mem{OV} < 1.0$)  according to
\citet{freytag:96}. For deep envelope (or core) convection
$f_\mem{ov}$ can be expected to be considerably smaller since the
ratio of the Brunt-V\"ais\"al\"a time-scales of the stable to the unstable
layers decrease with increasing depth, i.e. adiabaticity
\citep{herwig:97}.

Relevant in this regard are the 2D simulations of interior O-burning
shell convection in massive stars immediately before a supernova
explosion of \citet{bazan:98} and \citet{asida:00}. In these
simulations hydrodynamic fluid motions are not confined to the
convectively unstable region. The authors describe the excitation of
gravity waves in the adjacent stable layers \citep{hurlburt:86}, as
well as an exponential decay of convective velocities at and beyond
the convective border. As a result highly reactive fuel in the form of
C is mixed from above into the O-shell. However, the authors caution
that the quantitative results may suffer some uncertainty due to
numerical effects like resolution. The main difference between these
simulations and our regime of the He-shell flash is the lower driving
energy in our case, and we expect correspondingly lower Mach numbers,
less mixing and less overshooting.  \citet{young:05b} present initial
results on updated O-shell 2D and 3D convection simulations and apply
their findings on hydrodynamics mixing to 1D stellar evolution models
of supernova progenitors \citep{young:05a}. The internal structure in
these models is markedly different from standard models, with
important implication for the supernova explosion properties.

\subsection{About this paper}
The main purpose of our study is to obtain a characterization of the
hydrodynamic properties of shell convection in low-mass stars which is
determined by weaker driving due to smaller nuclear energy generation
compared to massive stars. We investigate the typical morphology, the
dominating structures and the time-scales of shell-flash convection,
and how these properties depend on resolution and driving energy. We
identify areas of parameter space that require further study in two-
and three-dimensions. In our present work we do not intend to quantify
mixing across the convective boundary, but we want to gain insight for
future simulations that shed more light on this important question.

In the following section we describe the 1D stellar evolution code and
the hydrodynamics codes used in this work. \kap{kap:setup} describes
the stellar evolution sequence and the specific model that serves as a
template for the hydrodynamic simulations. Then we describe the
results in \kap{kap:results}; specifically we describe the general
properties of the convection, the onset of convection, the evolved
convection and the convective and oscillation modes observed in our
simulations. We close the paper with a discussion (\kap{sec:concl}) of
the results, in particular the comparison of the 1D MLT velocity
profile and the time-averaged vertical velocity profile from our
hydrodynamic simulations.

\section{Codes}
\label{kap:codes}

\subsection{Stellar evolution code}
Our 1D hydrostatic, Lagrangian stellar evolution code EVOL with
adaptive mesh refinement and time stepping is equipped with up-to-date
input physics \citep[][and references therein]{herwig:03c}. Mass loss is
included according to the formula given by \cite{bloecker:95a} with a
scaling factor $\eta_\mem{BL}=0.1$. Nucleosynthesis is considered by
solving a sufficiently detailed nuclear network, with reaction
rates from the NACRE compilation \citep{angulo:99}. We use OPAL
opacities \citep{iglesias:96} and low-temperature opacities from
\citet{alexander:94}. In the stellar evolution code convective energy
transport is described by the mixing-length theory
\citep{boehm-vitense:58} with a constant mixing length parameter of
$\alpha_\mem{MLT} = 1.7$ determined by calibrating a solar model
with the parameters of the sun. For material mixing we solve a
diffusion equation for each species with a convective diffusion
coefficient as given by \citet{langer:85}.

\subsection{Hydrodynamics code}
For the hydrodynamic simulations we use the multi-dimensional
radiation-hydrodynamics code RAGE (Radiation Adaptive Grid Eulerian)
designed to model a variety of multi-material flows
\citep{baltrusaitis:96}. The conservation equations for mass,
momentum, and total energy are solved through a second-order,
direct-Eulerian Godunov method on a finite volume mesh
\citep{dendy:05}.  RAGE has been extensively tested on verification
problems \citep{belle:05,holmes:99,hueckstaedt:05}. The code has the
capability for continuous adaptive-mesh refinement to increase spatial
resolution in areas of interest with a minimal increase in
computational time. This capability is in particular useful for
multi-fluid applications. We do not use this feature because all our
runs have a single-fluid.



\section{Stellar evolution models of the He-shell flash and initial multi-dimensional setup}
\label{kap:setup}

We simulate the He-shell convection zone at a time point during a
thermal pulse evolution immediately before the peak of the
He-luminosity. At this time the He-burning luminosity that drives the
convection is already large implying a rapid build-up of an entropy
excess that drives convection. At the same time the layer has not yet
been expanded too much, and the number of pressure scale heights
involved is not excessive. Nevertheless, our simulations vertically
span $11.4\mem{H_p}$. Our inital conditions are a a set of polytropic
profiles that closely resembles the structure from the full 1D stellar
evolution model.

\subsection{The stellar evolution model}
\label{kap:evol-model} 

Our hydrodynamic simulations are based on a model from run ET2 of
\citet{herwig:04b} that represents typical properties of a thermal
pule in a low-mass asymptotic giant branchn star. This sequence has
metallicity $Z=0.01$ and the main-sequence initial mass is $2\msun$.
The evolution track was followed from the pre-main sequence until all
envelope mass was lost at the tip of the AGB. We assumed exponential,
time- and depth-dependent overshooting at the bottom of the convective
envelope ($f_\mem{ov}=0.016$). At the bottom of the He-shell flash
convection zone a very small overshooting ($f_\mem{ov}=0.001$) was
applied, mainly for numerical reasons. Such a small overshooting value
has little noticeable effect on the evolution but improves the
convergence of the code. The diffusion coefficient drops 5 orders of
magnitude within only 7 grid points corresponding to about $7
\mem{km}$, or less than $0.01\mem{H_p}$. Sequence ET2 has 16 thermal
pulses, eight of which are followed by a third dredge-up.

We choose a model from the second to last ($15^\mem{th}$) thermal pulse of the ET2
run. At this thermal pulse, the core mass
is $0.6\msun$ and the entire stellar mass is $1.4\msun$. For a stellar
model as template for the initial stratification of the hydrodynamic
simulations, we favor a large luminosity which implies large convective
velocities. However, the large luminosity quickly leads to a
significant expansion of the layers above the He-shell (\abb{fig:t-r}),
which dramatically increases the number of pressure scale heights that
the simulation would have to cover (\abb{fig:t-lhe-p}). In particular, we
would like to include enough stable layer above and below the
convection zone to get an impression of the hydrodynamic fluid
behavior at the convective boundaries and the adjacent stable regions.

We picked model 70238 of that rerun as a template for our hydrodynamic
initial setup. This model is one time-step ($0.07\jahre$) before the
peak of the He-shell flash luminosity of $4.29 \cdot 10^7\lsun$ (model
70239). The thermodynamic and mixing properties of model 70238 are
given in \tab{tab:prop70238} for the top and bottom of the convection
zone and the location of where we chose the top and bottom of the
hydrodynamic simulation box to be. The simulations covers a total
$11\, \mem{H_p}$, roughly half of this inside the convection zone.

\begin{figure}
\plotone{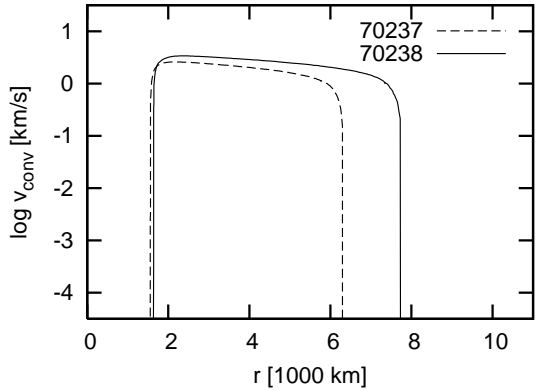} 
\figcaption{\label{fig:r-vconv} The velocity profiles from two
subsequent models of the 1D stellar evolution model track using the
mixing-length theory. Our hydrodynamic simulations resemble the conditions of model 70238. }
\end{figure}

The molecular weight profile and the convective velocity $v_\mem{conv}=D /
(\frac{1}{3} \alpha_\mem{MLT} H_\mem{P})$ are shown for two
consecutive models (70237 and 70238) in \abb{fig:r-mu} and
\ref{fig:r-vconv}. The time step between the two models is $0.04\jahre$
corresponding to approximately $1400$ convective turnovers. 
The radius integrated He-burning energy in the two subsequent models
$70237$ and $70238$ is $4.7\cdot10^{17}$ and $1.1\cdot10^{18}\mem{cm\,
  erg/g/s}$. During the He-shell flash, the nuclear energy is mainly generated by
the triple-$\alpha$ reaction.

\subsection{The hydrodynamic simulation setup and run parameters}
\label{kap:multi-D-setup}

We simulate a plane-parallel layer in Cartesian coordinates which
represents the intershell of the AGB star at the time nearly at the
peak He-shell flash luminosity.  In our models (lc0 for 2D and lc1 for
3D), the grid is equidistant in all dimensions, with a cell size of
$55\mem{km}$ in our standard mesh g (600x200). The box covers
$33000\mem{km}$ horizontally and $11000\mem{km}$ vertically in the 2D
runs.

The bottom of the simulation box is located at a stellar model radius
of $7507.5\mem{km}$, and subsequently all length scales are with
reference to the box bottom.  The bottom of the convection zone is at
$1650\mem{km}$. We approximate the stellar evolution model
stratification in the simulation box by a combination of three
polytropic stratifications \citep{hurlburt:86}, with $\gamma$ in each layer approximating
the $\gamma$ in the corresponding layer in the stellar evolution
model. For the region below the convection zone we assume a stable
polytropic stratification with $\gamma = 1.2$. For the convective
region we set $\gamma=\gamma_{\rm ad} = 5/3$, while the top polytropic stable
stratification mimics the stellar model with $\gamma = 1.01$.  At the
bottom of the convection zone we set the temperature to
$2.48\cdot10^8\kelv$ and the density to $1.174\cdot10^4\mem{g/cm^3}$
as in the stellar model(see \abb{fig:qinit_xc3}).
We fix the temperature below the convection
zone and at the top of the lower stable layer at $1.17\cdot10^8\kelv$,
corresponding to a factor of $0.47$ between the peak temperature at the bottom
of the convection zone and the local temperature minimum just below the
convection zone. The location of the top of the convection zone is set
to $7732.4\mem{km}$ from the bottom of the box, corresponding exactly
to the value in the stellar model. The top of the box at
$11000\mem{km}$ from the box bottom is chosen to be just below the
H-rich envelope. \abb{fig:qinit_xc3} demonstrates the agreement between our piecewise polytropic
stratification and the stellar evolution template model.

In the RAGE code the molecular weight of a material is specified in
terms of the specific heat, which for an ideal gas  is
\begin{math}
c_\mem{V} = \frac{3}{2}\frac{\mathcal{R}}{\mu} \komma
\end{math}
with the universal gas constant $\mathcal{R} = 8.31\cdot10^{7}
\mem{erg/K/g} $. $\mu-$gradients are present at the bottom and top of
the convection zone (\abb{fig:r-mu}). Although they add stability and
reduce mixing across the boundaries, we ignore $\mu-$gradients in this
set of calculations. We plan to address the effect of $\mu$-gradients in a
future study.  Here, we chose as a single mean molecular weight
$\mu = 1.4$ which closely matches the value in the convection zone
from the stellar model. Finally, we assume a constant gravity of $\log
g =7.7 \mem{[cgs]}$ vertically directed downward as in the convection
zone of the stellar model. In the stellar model gravity varies from
$\log g = 8.05$ at the bottom of the simulated layer to $\log g =
7.35$ at the top.

In the stellar evolution model, the nuclear energy that drives the
convection is calculated by solving a nuclear network based on
realistic temperature-dependent nuclear reaction rate input. The same
approach used in the hydrodynamic calculation may introduce additional
resolution dependencies of localized flow morphologies, in particular
at the bottom of the convection zone. We assume here a time-,
temperature- and resolution-independent volume heating in a small
layer at the bottom of the convection zone that releases the same
integrated amount of energy as the stellar evolution model.
Specifically, we release approximately $\epsilon_0 =
2\cdot10^{10}\mem{erg/g/s}$ over $550\mem{km}$ at the bottom of the
adiabatic stratification. A constant volume heating is realized by
multiplying $\epsilon_0$ with the density at the bottom of the
convection zone for the input value ($\epsilon^\star = \epsilon_0
\rho_0$) and then locally in each cell adding $\epsilon =
\epsilon^\star / \rho$ to the cell energy.

We introduce random density and temperature perturbations in pressure
equilibrium on the grid-level of $\log \Delta T/T = -4$ and centered
at the unperturbed values. Due to the discretization error, the
initial numerical stratification slightly violates hydrostatic
equilibrium.  This would cause a spectrum of plane-parallel p-mode
oscillations with significant velocity amplitudes. During an initial
damping phase ($700\mem{s}$) of the simulations we damp those
artificial p-modes to amplitudes at the $\mem{cm/s}$-level. The
damping is achieved by multiplying a factor to the momentum in each
cell. This damping factor decreases from an initial value to zero over
the damping phase. A corresponding heating factor increases from 0 to
1 over the damping phase and smoothly turns heating for convection
driving on. The length of the damping phase was chosen so that an
overall velocity minimum could be achieved before the onset of
convective motions. 

We formulated a grid of runs in 2D that cover a range in resolution
and heating rate (\abb{fig:grid}).  These runs are labeled lc0AB,
where lc = luminosity driven convection, 0 is the run series, in this
case a series of 2D runs, A denotes the heating rate and B the grid
size.  We have done one 3D run, denoted by lc1. We also discuss two
runs (lc2) in which we have used three formally different species in
the three layers of our setup, but the three materials have the same
material properties. This allows some preliminary insight into mixing.

\subsection{Time-scales and summary}
The convective turn-over time scale is of the order of $600\mem{s}$
and convective velocities are of the order a few $\mem{km/s}$. We use
an explicit, compressible hydrodynamics code, in which the time step
is limited by the Courant-Friedrichs-Levy condition imposed by the
sound speed. The maximum sound speed in our setup is
$>1000\mem{km/s}$.  Accordingly, our convective flows have a low Mach
number of the order $10^{-3}\mem{Ma}$. With a grid size of
$55\mem{km}$ this implies a time step of $\Delta t <
5\cdot10^{-2}\mem{s}$. In order to achieve some convective
steady-state and evolve this configuration for some time to have good
statistics, our multi-dimensional simulations run for approximately 15
to 30 convective turn-over times, which corresponds to more than
$10^5$ time steps. Thus, the multi-dimensional simulation time is much
less than $0.1\%$ of the stellar evolution calculation time-step that
reflects the thermal time scale. Our calculation represent a snapshot
in time compared to the long-term evolution of the entire He-shell
flash which takes $200 - 300\jahre$. In our simulations we assume
a constant heating rate in time. For the highly adiabatic He-shell
flash convection the ratio between convective and non-convective heat
transport (the Peclet number) is very large. For that reason we ignore
thermal transport in this set of simulations.

By running a series of simulations with heating rates spanning a
factor $1000$ we study how hydrodynamic properties depend on the
heating rate. We also want to derive how our results depend on
resolution. \abb{fig:grid} gives an overview of our 2D runs for
our standard single-fluid setup lc0.

\section{Results}
\label{kap:results}

Our simulations start with a Raleigh-Taylor like growth
sequence of the initial perturbations driven by the heating at the
bottom of the unstable layer (\kap{sec:onset}). Eventually, fully
developed convection emerges (\kap{sec:evolconv}). It features a
cellular flow configuration best seen in the pseudo-streamline plot in
\abb{fig:lc0gh_prelfluct_single}.  The flow field is dominated by a
few large cells (about 4 in that example) each consisting of a
clockwise and a counter-clockwise vortex. Vertically the cells span
the entire convectively unstable region. However, the vortex centers
are located closer to the bottom, and leave room for some substructure
in the top layers of the convectively unstable zone.  Additional
smaller rolls also appear close to the bottom boundary.  Gravity waves
are excited in the stable layers. An oscillaton mode analysis of the
their appearance at the transition from unstable to stable
stratification can be used to better understand the interaction of
stable and unstable layers (\kap{sec:modes}).
\abb{fig:lc0gh_sfluct_single} shows the corresponding entropy
fluctuations for the same snapshot.  Hot material rises from the
heating layer at the bottom of the convection zone. The bulk of the
high-entropy areas starts to turn side-ways well below the top of the
unstable layer. However vertical high-velocity streams persist through
the unstable layer, and the velocity field can be perpendicular to the
streaks of high-entropy material.

\subsection{Onset of convection}
\label{sec:onset}

The drag force applied in the first 700\,\mem{s} (p-mode filter,
\kap{kap:multi-D-setup}) of a simulation eliminates efficiently all
modes that appear due to the imperfect hydrostatic equilibrium of the
initial stratification.  On the other hand, it also delays the onset
of convection somewhat.

After some time, the external heating builds up a small extra entropy
bump at the bottom of the entropy plateau.  The top of this bump is
the place where the entropy has the steepest negative slope, and where
the convective instability causes small-scale modes to grow first (see
the top panel in Fig.~\ref{fig:lc0gh_sfluct_onset}).  The modes expand
vertically and the velocity amplitude increases.  Later, when they
leave the linear regime, they form a patch of tiny mushroom-like
instabilities that are similar in their morphology to the
Raleigh-Taylor instability.  These grow and merge further forming
complex large-scale structures (Fig.~\ref{fig:lc0gh_sfluct_onset}).

The stable regions are far from being quiet but instead filled with
waves.  Remarkably, these waves occur in the upper stable zone well
before the convective plumes reach this layer.  The motions are not
excited by plumes ``hitting'' the boundary layer but by the pressure
excess in front of the plumes. This can be clearly seen in
\abb{fig:multiplotfield_lc0gh_onset_P-relfluct_vstream} which shows
the pressure fluctuations and pseudo-streamlines of the same time and
model as the middle panel of \abb{fig:lc0gh_sfluct_onset}. The rising
plumes have not yet reached the top of the unstable layers; however
the entropy fluctuations at and above the top of the unstable layer
show that oscillation modes are already excited by the fluctuating
over-pressure regions seen, for example, at the horizontal position
$x$=25\,\mem{Mm} in
\abb{fig:multiplotfield_lc0gh_onset_P-relfluct_vstream}.  We
demonstrate in Sect.~\ref{sec:modes} that these oscillation modes are
in fact gravity waves.

Figure~\ref{fig:lc0dX_sfluct_researly} demonstrates that the amount of
detail visible in the plumes strongly depends on the resolution while
the number of plumes (or the size of the largest structures at a given
time-step) is only slightly affected.  The wavelength of the modes
that initially start to grow and also the size of the first generation
of mushrooms decreases with grid size.  However, these small-scale
structures quickly merge to form much larger cells.

We do not present models with a coarser grid than 210x70 because test
runs have shown that numerical problems arise at the upper boundary of
the box (an artificial entropy inversion and tiny ``surface cells'').
That is not particular worrisom; a hydrodynamics code needs a minimum
number of grid cells to resolve a pressure scale height properly, and
we have about 11.4\,\mem{H_p} inside our box.

In the low-resolution model (top panel in
Fig.~\ref{fig:lc0dX_sfluct_researly}) the entropy above the
``mushrooms'' is not constant as it should be for perfectly adiabatic
flows. This artifact essentially vanishes with increasing resolution
and does not seem to have any adverse effect.

To demonstrate the influence of the heating rate upon the convective
patterns during the onset phase, a scaling of the amplitude of the
entropy inhomogeneities (the fluctuations increase with the heating)
and adjustment of the selected time-steps are necessary (a larger
heating rate causes larger convective velocities and shorter growth
times, see Fig.~\ref{fig:lc0Xg_sfluct_heatearly}).  Taking the scaling
into account, the resulting morphology of the convective flows are
remarkably similar, suggesting the possibility of avoiding excessively
CPU-intensive simulations with low heating rates and correspondingly
small convective velocities.

Careful inspection of the lower panels of Fig.~\ref{fig:lc0Xg_sfluct_heatearly}
reveals spurious entropy minima as dark spots at the top of the entropy plateau
caused by a small negative artificial entropy spike at that position.
While the employed initial p-mode filter facilitates the relaxation
to a complete numerical hydrostatic equilibrium,
there is also a thermal relaxation going on with  a much longer
time-scale (hundreds of seconds instead of seconds).
This process leads to a smearing of the initially sharp jump in
density and internal energy at the bottom of the entropy plateau
and in the discontinuity in the derivatives at the top of the plateau.
This redistribution of density and energy does not preserve the entropy profile
but causes an artificial (small) entropy maximum and minimum to occur at
the bottom and the top of the entropy plateau, respectively.
The amplitude and the width of the entropy spikes decrease with
increasing resolution.
However, at low resolutions and low (i.e.\ realistic) heating rates,
they cause an initial driving of the convective flow
comparable to the one due to the external heating.

Fortunately, this additional driving is a transient phenomenon because
there is only a limited amount of energy available that can be
``extracted'' by a conservative code out of the discontinuities.  It
only increases the time it takes for the convection to settle to a
statistically stationary state by approximately $3000\sec$ (for the
run lc0gg) depending on resolution.



\subsection{Onset of convection: 2D versus 3D}
\label{sec:dim}

\begin{figure}
  \plotone{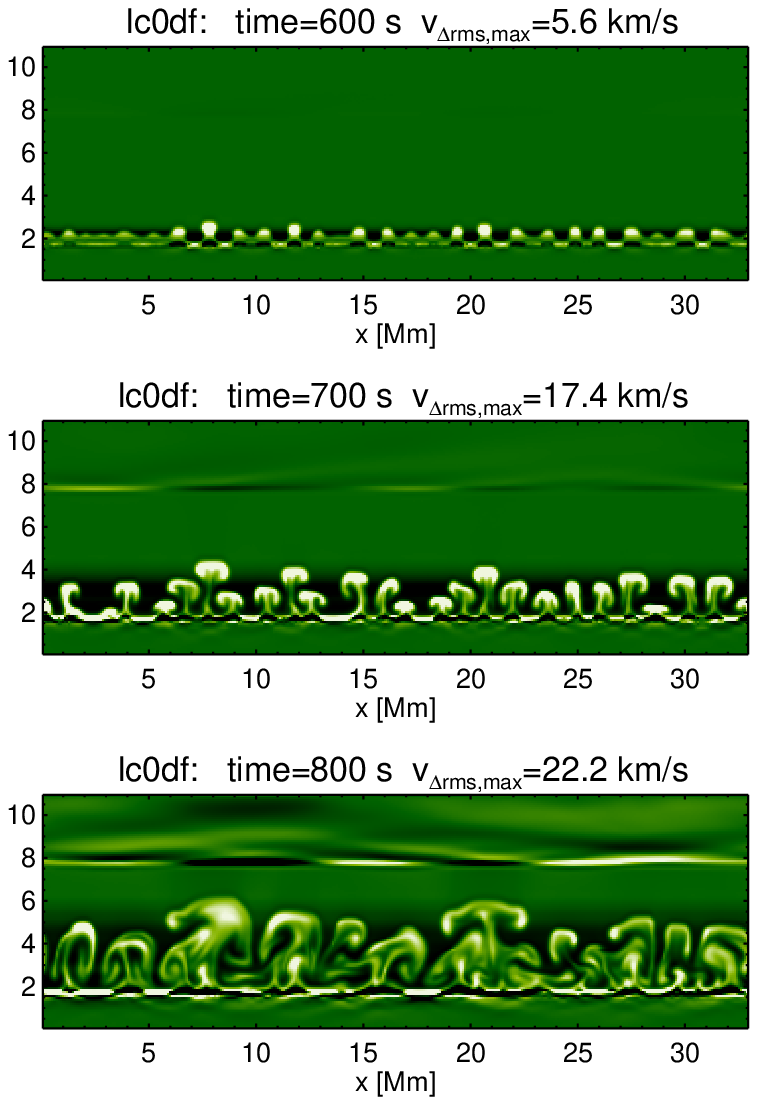}
  \figcaption{\label{fig:lc0df_sfluct_single} 2D lc0df run sequence of
    entropy inhomogeneities during the onset of convection.
    The same snapshot times and scaling is used as in
    Fig.~\ref{fig:lc1df_sfluct_single}.}
\end{figure}

\begin{figure}
\plotone{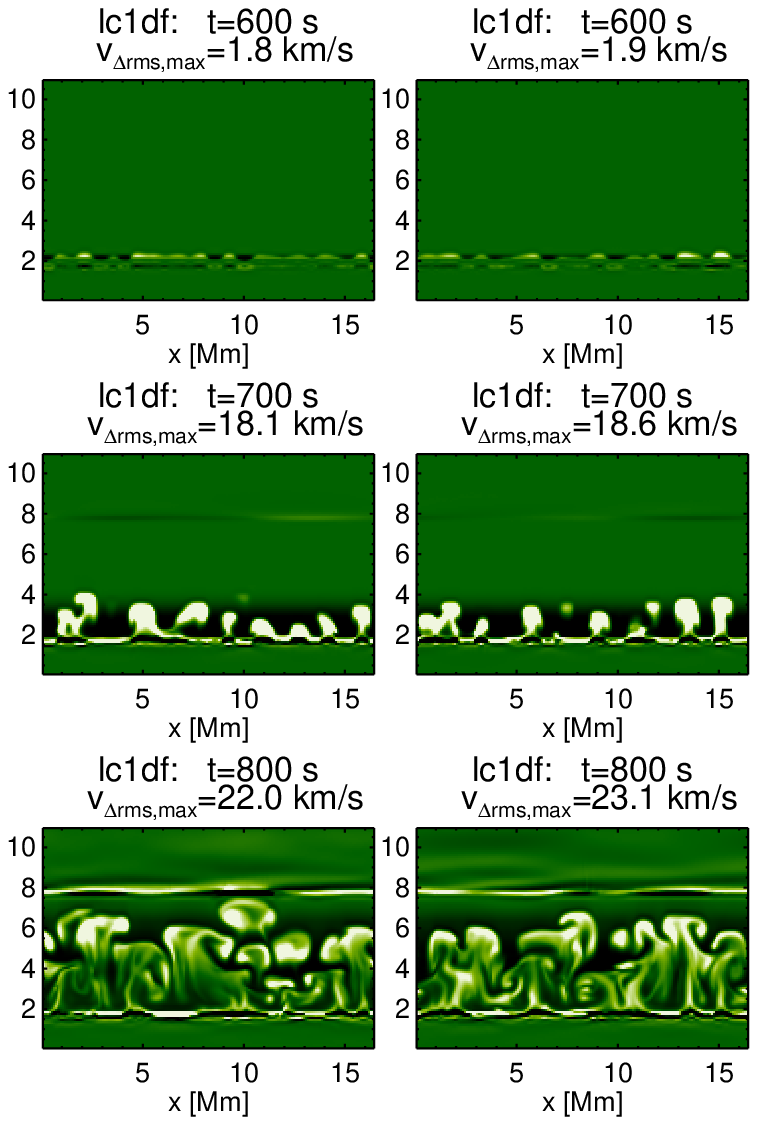}
\figcaption{\label{fig:lc1df_sfluct_single} 3D lc1df run sequence of
the entropy inhomogeneities during the onset of convection.
The same snapshot times and scaling is used as in Fig.~\ref{fig:lc0df_sfluct_single}.}
\end{figure}

In addition to the two-dimensional models, we have some information
from a single 3D run (lc1d, 1502x100 grid points).
Figs.~\ref{fig:lc0df_sfluct_single} and \ref{fig:lc1df_sfluct_single}
display entropy inhomogeneities during the onset of convection for the
2D and the 3D run, respectively.  Heating rate, vertical extent, and
grid cell size are the same in both models.

The 3D model initially has smaller velocities and appears to be
slightly behind the 2D model.  However, as soon as convection sets in,
the velocities in the corresponding snapshots are very similar.  The
structures in the last panels seems to be slightly smaller in 3D than
in 2D.  However, distorted ``mushrooms'' appear in both sequences.
Also, gravity waves are excited in both models.

We do not see any hint in the 3D models that would indicate that the
results we derive from the grid of 2D models would suffer from the
restriction to two dimensions.  To get accurate quantitative results
3D models are to be preferred.  However, 3D models with high
resolution and realistically low heating rates are computationally
extremely demanding.  We will try to approach this regime in future
work.

\subsection{Evolved convection}
\label{sec:evolconv}

We demonstrate the properties of fully developed convection by means of
sequences of snapshots and averaged 1D profiles.

Figure~\ref{fig:lc0gh_sfluct_evol} shows an example of the evolution
of the convective patterns in a later stage of the 1200x400 run with
realistic heating rate.  During this time, the flow is dominated by 4
global cells (the aspect ratio of the unstable zone is about 6).  The
centers of the vortices are located in the lower half of this zone,
characteristic for convection in a strongly stratified environment, as
found by \citet{hurlburt:84} in their stationary convection.  However,
in our case the upflows, starting in the thin heating layer at the
bottom, are dynamic and change their shapes all the time.  Smaller
rolls can occur, especially at both sides of the foot-points of the
updrafts and also near the top of the entropy plateau.

The stability of the general pattern is supported by the fact
that the down-flows tend to inhibit the development of new updrafts
that can develop due to the convective instability in the heated
bottom layers.
Instead, the heated material is pushed sideways into the existing
upflow channels.
During this process, new upflow ``fingers'' can develop.
Most often, they are washed into the larger updrafts.
However occasionally, they have a chance to grow and to change
the global pattern.

The situation is pretty much the inverse of the well-known scenario of
stellar surface convection, where radiative cooling creates a thin
super-adiabatic layer that leads to the formation of cool downdrafts
that drive the convection.  Whereas in surface convection the
stratification tends to focus the downdrafts \citep{nordlund:97}, the
hot upflows expand.  This expansing effect can widen the thin columns
of hot upflowing material in the case of He-shell flash convection.
Nevertheless, they can rise over several pressure scale heights before
they break up and can traverse the entire convection zone.

\begin{figure}
  \plotone{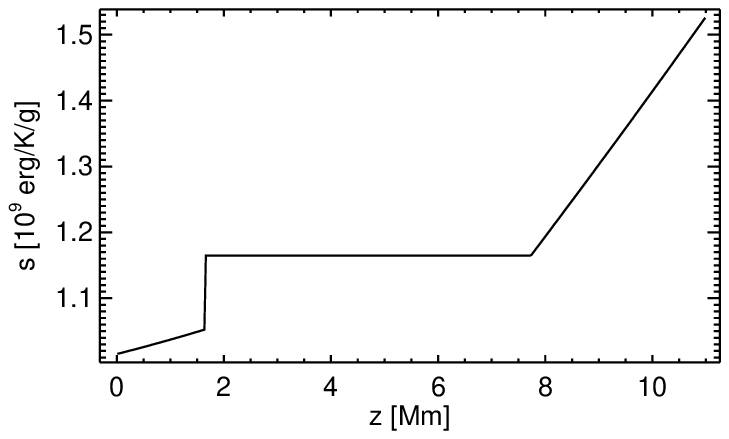}
  \plotone{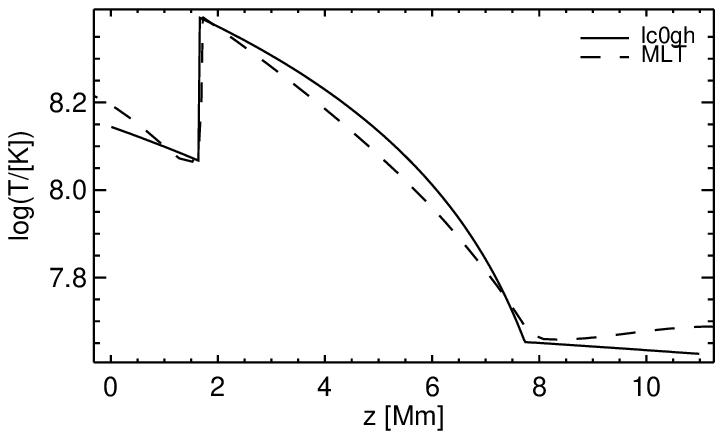}
  \plotone{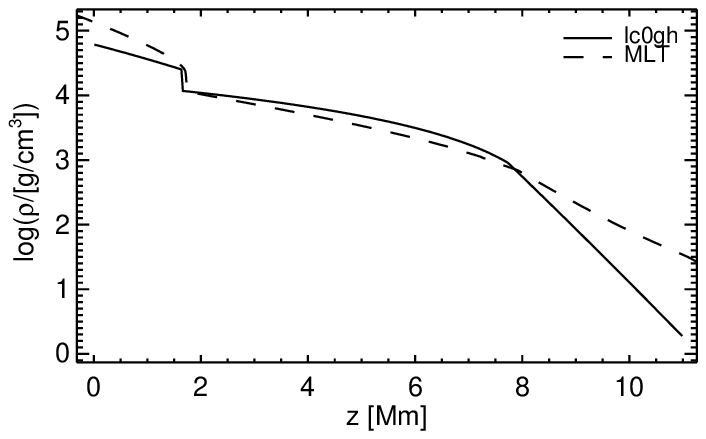}
  \plotone{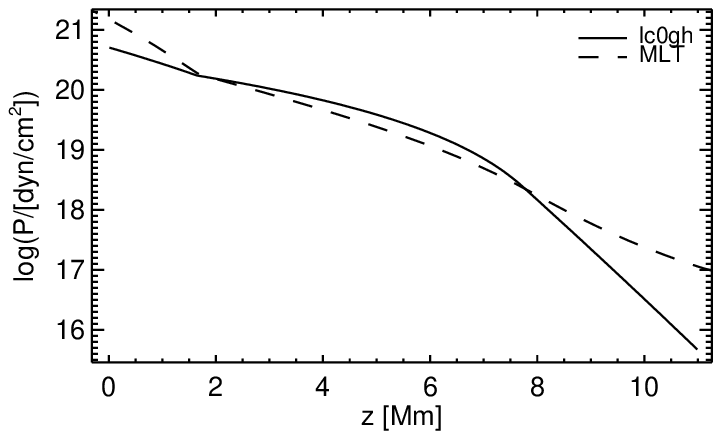}
  \figcaption{\label{fig:qinit_xc3} The initial entropy, temperature,
    density, and pressure stratifications of the hydrodynamic
    simulations (solid line) and the temperature, density and pressure
    from the 1D model ET2/70238 (dashed line).}
\end{figure}

The entropy jump at the bottom of the plateau and the sudden
increase at the top (see Fig.~\ref{fig:qinit_xc3})
confine the convective flow essentially
to the region in between -- with very narrow transition layers.
In these boundary layers the largest differences between
the runs with different resolutions and heating rates occur.
To compare them quantitatively, we compute some averaged quantities.

During the simulations a snapshot is stored every 0.5\,\mem{s}.
This results for each quantity $q$
(the velocity components, internal energy, $\rho$, $T$, $P$) 
in a data-set $q(x,y,t)$
depending on horizontal coordinate $x$,
vertical coordinate $y$,
and time $t$.
This output high rate clearly over-samples the slow convective motions
especially in the runs with low heating rates.
However, it is justified for the fastest gravity waves.
And it is just barely sufficient to sample the pressure waves
(see Sect.~\ref{sec:modes}).

For each snapshot we compute horizontal averages,
\begin{equation}
  q_{\rm avg}(y,t) = \langle q(x,y,t) \rangle_{x}
\end{equation}
and horizontal root-mean-square (rms) fluctuations
\begin{equation}
  q_{\rm \Delta rms}(y,t) = \left( \langle \left( q(x,y,t)-q_{\rm avg}(y,t) \right) ^2 \rangle_{x} \right)^{1/2}
    \enspace .
\end{equation}
We check the time evolution of these quantities to ensure that the
simulations have reached a statistical steady state. Due to heating at
the bottom of the convective layer the entropy of the convection zone
is slowly increasing and the extent of the convection zone is growing
at a small rate.  The time evolution of the maximum vertical
rms-velocity is shown for runs with standard and enhanced heating and
different resolutions in \abb{fig:t-vrms}. For the runs with the
standard heating rate g a steady-state according to this disgnostic is
reached after $6000\mem{s}$, following the onset phase of convection
which is characterized by larger velocities. At a higher heating rate
the difference between the initial velocity peak during the onset
phase and the following steady-state velocity is smaller. In this case
the steady state is already reached after approximately
$3000\mem{s}$.

\begin{figure*}

  \includegraphics[width=5.5cm]{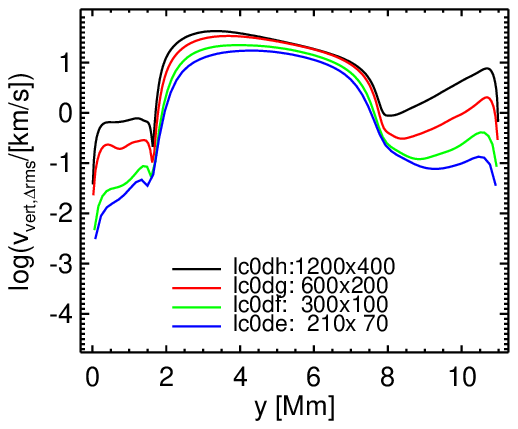}
  \includegraphics[width=5.5cm]{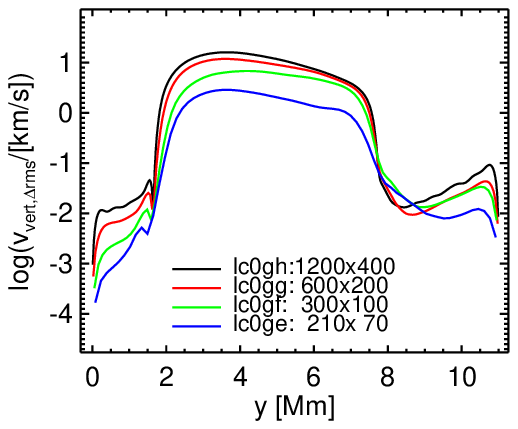}
  \includegraphics[width=5.5cm]{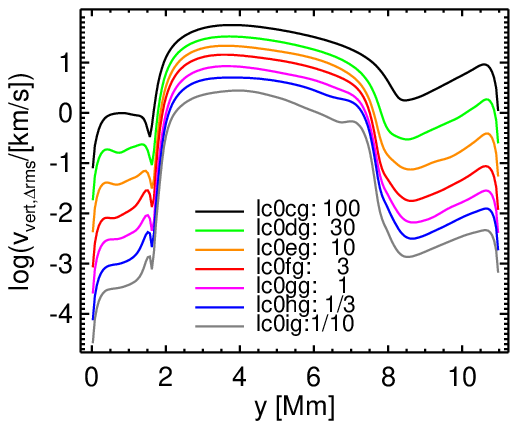}
  \includegraphics[width=5.5cm]{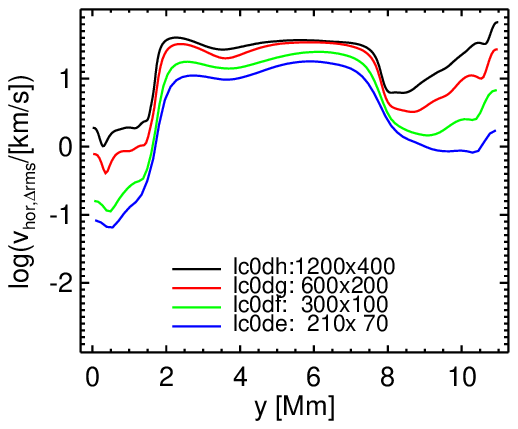}
  \includegraphics[width=5.5cm]{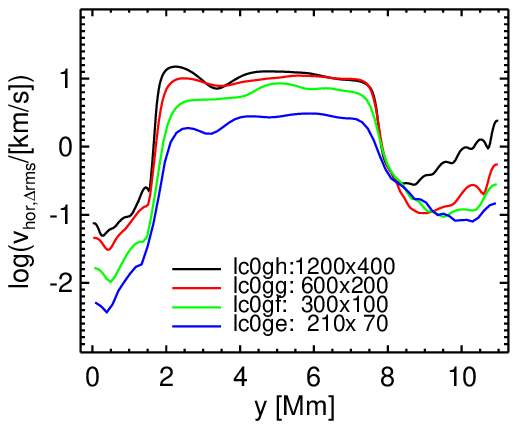}
  \includegraphics[width=5.5cm]{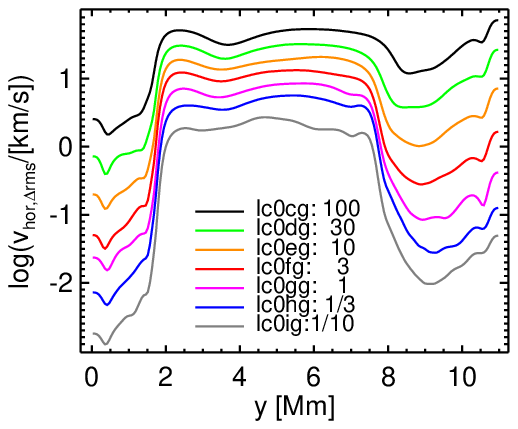}
  \includegraphics[width=5.5cm]{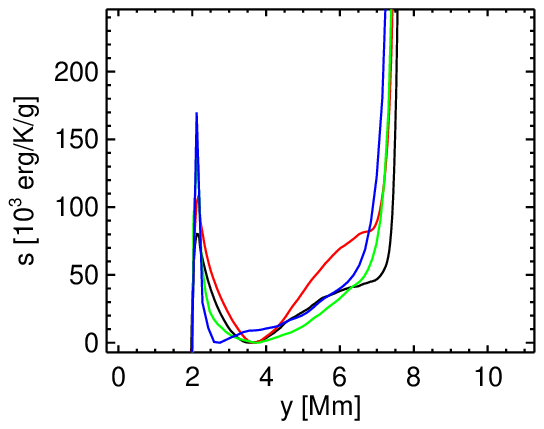}
  \includegraphics[width=5.5cm]{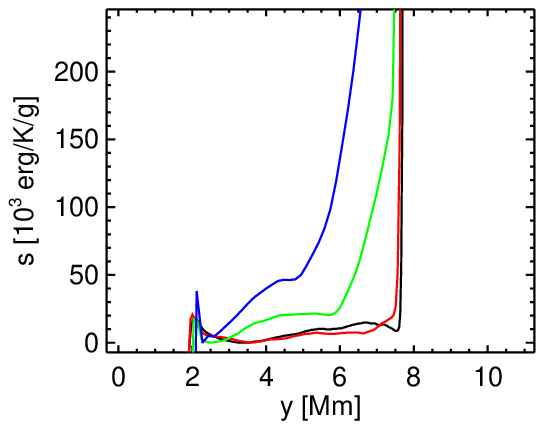}
  \includegraphics[width=5.5cm]{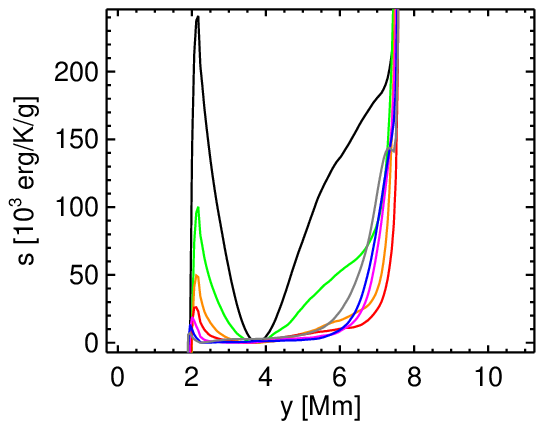}
  \includegraphics[width=5.5cm]{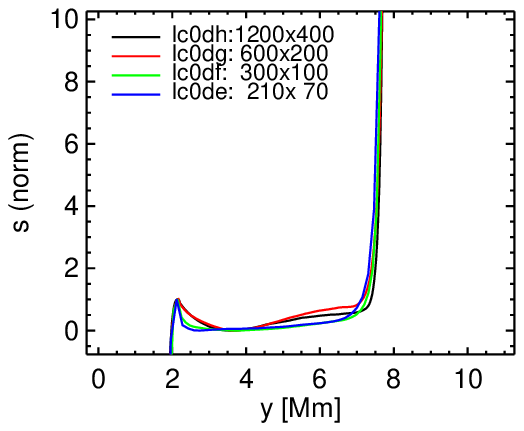}
  \includegraphics[width=5.5cm]{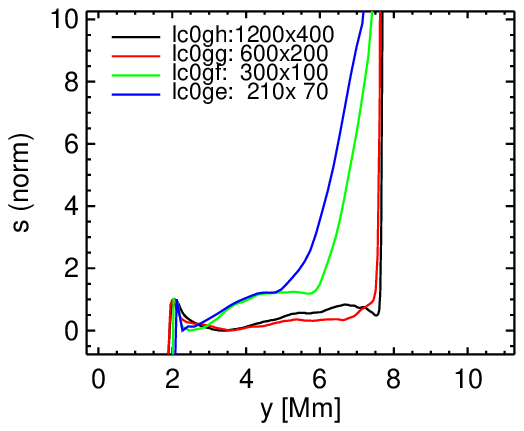}
  \includegraphics[width=5.5cm]{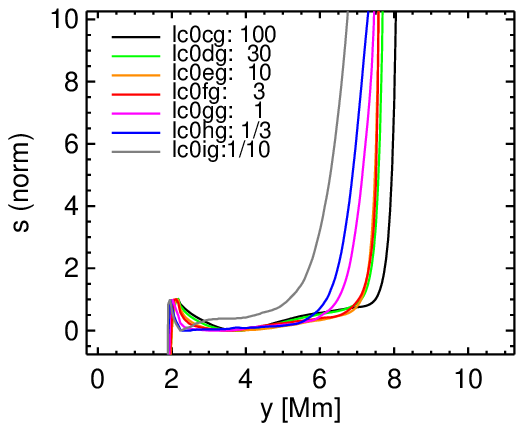}
  \figcaption{\label{fig:stat} Some time-averaged quantities for
    various model sequences.  Each column of panels refers to the same
    sequence.  All panels in one row show the same quantity.  Left
    column: runs with different resolutions and large heating rate
    (lc0dX).  The averages are performed over the interval
    1500-4500\,\mem{s}.  Center column: runs different resolutions and
    realistic heating rate (lc0gX).  The averages are performed over
    the (short) interval 3750-4300\,\mem{s}.  Right column: runs with
    intermediate resolution (600x200) and different heating rates
    (lc0Xg).  The averages are performed over the last 5000\,\mem{s}
    of each sequence.  Top row: rms-fluctuations of the vertical
    velocity.  Second row: rms-fluctuations of the horizontal
    velocity.  Third row: mean entropy $s$ shifted so that the minimum
    entropy on the plateau becomes zero.  Bottom row: mean entropy $s$
    shifted as above and scaled to set the amplitude of the lower
    entropy bump to unity.  }
\end{figure*}

To compare models with different resolutions or heating rates we average these
quantities over time to get time-independent vertical profiles, as
\begin{equation}
  q_{\rm avg}(y) = \langle q_{\rm avg}(y,t) \rangle_{t}
\end{equation}
or
\begin{equation}
  q_{\rm \Delta rms}(y) = \left( \langle q_{\rm \Delta rms}(y,t)^2 \rangle_{t} \right)^{1/2}
    \enspace .
\end{equation}
Examples for the rms-fluctuations of vertical and horizontal velocity
are shown in Fig.~\ref{fig:stat}, first and second row, respectively.
Plotting the entropy $s$ on the same scale as in \abb{fig:qinit_xc3}
would show no difference to the initial stratification.  Therefore,
only the entropy difference to the minimum plateau value is plotted in
the panels in the third row, on a highly magnified scale.  To allow
the comparison of models with very different heating rates (and
entropy fluctuations) in the last row the same data is rescaled to
give the entropy bump at the bottom of the plateau a value of unity.

The velocities are clearly not confined to the region that would have been
predicted by MLT
(between 1.7\,\mem{Mm} and 7.7\,\mem{Mm},
compare the top panels in Fig.~\ref{fig:stat} with Fig.~\ref{fig:r-vconv}).
Nevertheless, before reaching the boundaries of the entropy plateau
the vertical velocities drop exponentially by orders of magnitude.
Close to the boundaries there are strong rapidly declining
horizontal velocities leading to significant shear flows
(second row of panels in Fig.~\ref{fig:stat}).

Although the classical boundaries of a convection zone do not
actually confine the velocity field completely,
they still indicate where the flow changes its character:
large overturning convective motions, inside
and small wave-like motions outside
(cf.\ Fig.~\ref{fig:stat}
and e.g.\ Figs.~\ref{fig:lc0gh_prelfluct_single}
and \ref{fig:lc0gh_sfluct_single}).
The low-heating models in Fig.~\ref{fig:stat} (top right panel)
indicate the presence of a transition region where the velocities
decline rapidly inside the top stable regions before they rise
due to the waves.

The dependence of the velocity on the heating rate is rather simple:
it grows monotonically with heating rate. The waves in the stable
layers have larger amplitude than the velocities of the overturning
convective motions (right column in Fig.~\ref{fig:stat}). It is
important to emphasize that the large vertical velocities shown in the
stable layers of the top panels of \abb{fig:stat} do not necessarily
indicate effective mixing (see \kap{sec:mixing} for details).

The dependence on the resolution is somewhat similar.  For the
convective motions, there are  hints towards convergence of the vertical and horizontal rms-velocities at the
highest resolutions models (1200x400).  However, the wave amplitude in
the runs with large heating rate shows less convergence.  The
(logarithmic) step in the amplitude of the horizontal velocities
between the 1200x400 and the 600x200 run is smaller than between the
600x200 and the 300x100 run.  Models with even higher resolution are
desirable to check the convergence.  
The appearence of similiar quantitative properties in He-shell flash
convection appears to be robust; these properties are present at all
resolutions and heating rates, in 2D and in 3D.

A higher resolution does not affect the number of large cells
noticeably.  However, it leads to finer entropy filaments and
complex sub-structures of the largest plumes (cf.\ 
Figs.~\ref{fig:lc0dX_sfluct_reslate} and
\ref{fig:lc0gX_sfluct_reslate}).  Less numerical dissipation on a
given scale leads to higher velocities, stronger velocity gradients
(causing more dissipation), and some additional smaller rolls.
However, even the high-resolution model in
Fig.~\ref{fig:lc0gh_prelfluct_single} shows the dominance of the
large-scale structures.  The trend to higher convective velocities is
limited by the amount of energy that has to be transported.

The situation is different for the velocity amplitudes of the g-modes
in the stable layers. A wave with a certain wavelength directly
feels the lower damping rate at a higher numerical resolution.
Since the wave excitation rate is larger due to the larger
convective velocities, a mode reaches a significantly larger
amplitude. This explains why the g-mode velocity amplitudes increase
with resolution.

\subsection{Detailed shape of the entropy plateau}
\label{sec:entplateau}

The initially flat entropy profile on the plateau changes during a simulation.
However, it does not show a simple monotonic decline as in a corresponding
MLT model.
Instead, it has a complex height and parameter dependence
(see the panels in the lower half of Fig.~\ref{fig:stat}).

All models show a sharp entropy peak in the heating zone at the
bottom of the entropy plateau.
The entropy declines with height, indicating a convectively unstable region
consistent with MLT.
However, the entropy minimum is not located at the top of the plateau
but somewhere in the middle, typically in the lower half.
Above the minimum, the entropy rises and has, in most models,
a smooth monotonic transition to the stable layers above the convection zone.
An exception is the model lc0gh
(with realistic heating rate and highest resolution,
cf.\ the two bottom panels in the center column in Fig.~\ref{fig:stat}):
It shows a second minimum at the top of the plateau.

Note, that even if the entropy gradient is positive in some parts
of the convection zone and therefore the stratification is formally stable there,
the matter is heated in the shallow zone at the bottom and the
plumes maintain their entropy excess and remain buoyant
over the entire convection zone
(indicated by the bright color of the plumes in
e.g.\ Fig.\ref{fig:lc0gh_sfluct_single}).

We interpret the entropy rise at the top of the plateau of
most models as due to mixing of high-entropy material from
the stable layers above into the convection zone.
This mixing depends on the velocity amplitude, the numerical
resolution, and time.

A significant part of the flow close to the upper boundary of the
entropy plateau is due to wave motions. These are almost reversible
and have a low mixing efficiency and cause a diffusion of entropy due
to numerical viscosity. In addition ro numerical mixing there may be a
finite amount of physical mixing.

While higher resolution means less dissipation it also means larger
amplitude of the motions that cause the mixing.  A large heating rate
means strong mixing but also a quick rise of the plateau so that the
relative effect remains small.  The model lc0cg (which has the largest
heating rate) has the largest absolute amount of mixing (the highest
entropy rise at upper part of the plateau, see third panel in right
column in Fig.~\ref{fig:stat}).  However, the relative effect is
smaller than for the models with lower heating rate (fourth panel in
right column in Fig.~\ref{fig:stat}).

The resolution effect onto the models with large heating rate (left
column in Fig.~\ref{fig:stat}) is only moderate.  The relatively quick
rise of the entropy plateau value seems to limit the region with the
entropy contamination from above.  The dissipation of entropy is
similar for all four tested resolutions.

In the models with realistic heating rate (center column in
Fig.~\ref{fig:stat}) the entropy profile changes qualitatively and
quantitatively.  The low-resolution models show a strong rise of
entropy, whereas the highest resolution model has a well-defined
minimum close to the top of the plateau.  The effect of this minimum
can be seen e.g.\ in Fig.\ref{fig:lc0gh_sfluct_single} or
Fig.\ref{fig:lc0gh_sfluct_evol} as tiny dark plumes emanating from the
top unstable layers downward.  An initial extra minimum in the entropy
profile of model lc0gg (600x200 points) vanishes during the
simulation.  The minimum of model lc0gh (1200x400) remains during the
available simulation time.  The profile for very high resolutions and
long simulation time is therefore not predictable, yet.

The change in the entropy profile due to this slow mixing has an
impact on the flow for the runs with low resolution and low heating
rate.  While the flow patterns during the onset and early stages of
convection are very similar when a proper scaling of the amplitudes of
the fluctuations is taken into account (see
Fig.~\ref{fig:lc0Xg_sfluct_heatearly}) the patterns differ after a
sufficiently long simulation time (see
Figs.~\ref{fig:lc0Xg_sfluct_heatlate} and
\ref{fig:lc0Xg_pfluct_heatlate}).  The contamination of the convection
zone with high-entropy material is so strong that the stratification
is partly stabilized.  The convective velocities in the top part of
the entropy plateau decrease, and the typical cell sizes become
noticeably smaller.

\subsection{Modes}
\label{sec:modes}

\begin{figure}
\includegraphics[width=7.6cm]{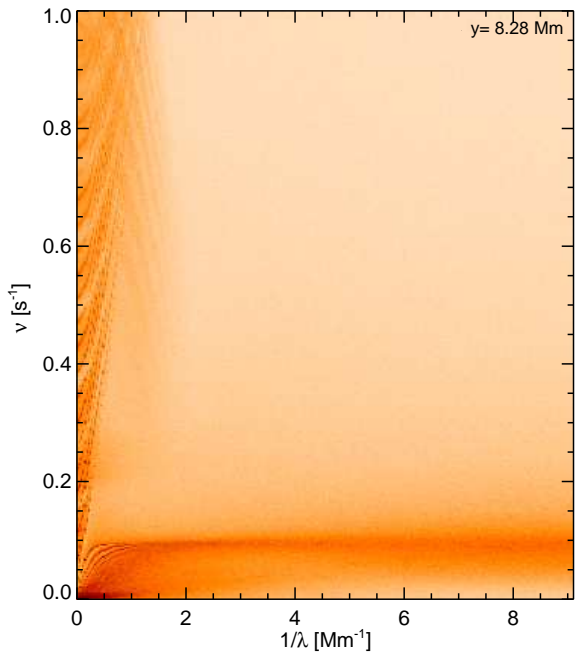}
\figcaption{\label{fig:komega_lc0gg_ex} $k$-$\omega$ diagram
(actually $1/\lambda$-$\nu$ diagram)
for an arbitrary height in the upper stable region for run lc0gg.
Both axes are linear and extend to the respective Nyquist frequencies.
They are given by the sampling rate in time,
$\nu_{t,\rm Nyquist}$=$1/2\Delta t_{\rm sampling}$=1\,s$^{-1}$
and the horizontal grid size,
$\nu_{x,\rm Nyquist}$=$1/2\Delta x_{\rm grid}$=9.09\,Mm$^{-1}$.
The signal strength goes from bright (low amplitude) to dark (high amplitude).
}
\end{figure}

To further investigate the waves encountered in the simulations, we
construct $k$-$\omega$ diagrams from Fourier analyses.  An example is
shown in Fig.~\ref{fig:komega_lc0gg_ex} for an arbitrary height point
inside the upper stable region for the run lc0gg.

In order to produce such a diagram, we collect all points for a single
quantity (the vertical velocity) for a chosen height and all time
steps of an interval into a 2D array.  For the example from the lc0gg
run, the array has a space axis with 600 points spanning 33\,\mem{Mm}, and a
time axis with 12000 points spanning 6000\,\mem{s}.  Next, a possible
trend in time is removed by subtracting a parabola fitted to the
time-dependent spatial average of the data.  This is actually not
necessary for the vertical velocities.  But it is essential for data
like the entropy that has a large average value and a trend in time.
To avoid signals due to the misfit of the beginning and end of the
sequence, the data is faded in and out by multiplying with a
cosine-bell function.  A 2D Fourier analysis is applied and the
power-spectrum is computed.  The latter is smoothed in frequency
space by applying a (1/4, 1/2, 1/4) filter.  To emphasize small
signals, the power$^{-0.1}$ is plotted in the examples.

A simple mode with a certain wavelength and period would give a single
(dark) spot in a $k$-$\omega$ diagram.  Imperfections (e.g.\ due to
interaction with convection or damping) would cause a smearing in
frequency.  Plane-parallel modes are found on the $\nu$ axis.  Static
signals (e.g.\ processes on time-scales exceeding the interval) are
located on the $1/\lambda$ axis.

While a horizontally moving sine wave would give a single spot, a more
complex function that travels horizontally with a certain speed would
produce an entire ray with a slope $\nu \lambda$.  Mode families give
rise to ridges. Their detailed shape depends on the background
stratification.  Convection does not consist of discrete modes.
Instead, it has only typical time and length scales and gives a
smeared blob.

On the left side of Fig.~\ref{fig:komega_lc0gg_ex},
pressure waves (p-modes) are visible, reaching up to the top
of the diagram.
In fact, a reflection of the pattern occurs at the top of the figure.
The sampling rate is not quite high enough to fully resolve all p-modes,
which leads to this aliasing phenomenon.

The bottom part of the diagram
(below the Brunt-V{\"a}is{\"a}l{\"a}-frequency of 0.1\,\mem{s^{-1}})
is dominated by gravity waves (g-modes).
At longer horizontal wavelengths (small $1/\lambda$) individual ridges are visible.
The top ridge (highest frequency) corresponds to the mode with largest
vertical wavelength.
Modes with smaller wavelength have a lower frequency
(contrary to the behavior of p-modes).
The g-modes with very small horizontal or vertical wavelength are
not resolved individually and only give a dark ``band''.

\begin{figure*}
\includegraphics[width=4.83cm]{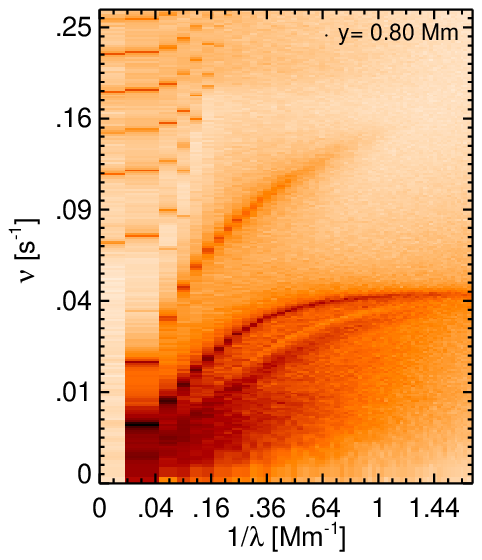}
\includegraphics[width=3.8cm]{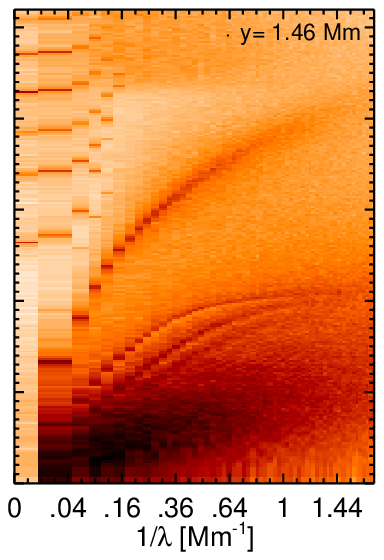}
\includegraphics[width=3.8cm]{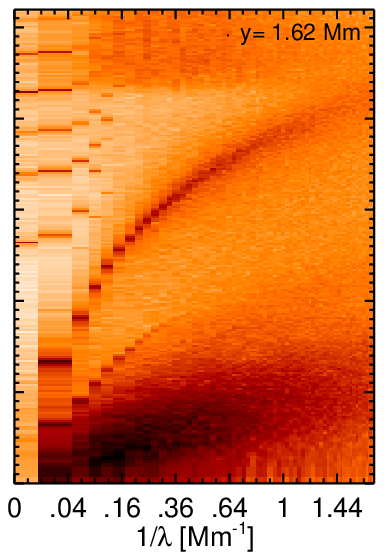}
\includegraphics[width=3.8cm]{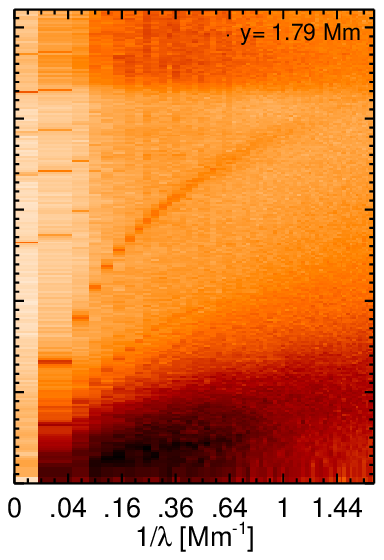}
\includegraphics[width=4.83cm]{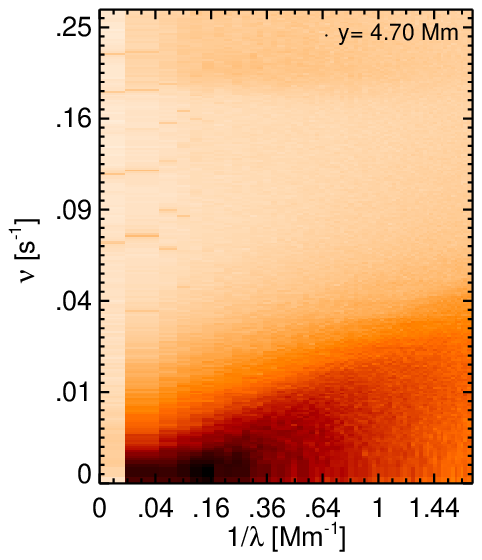}
\includegraphics[width=3.8cm]{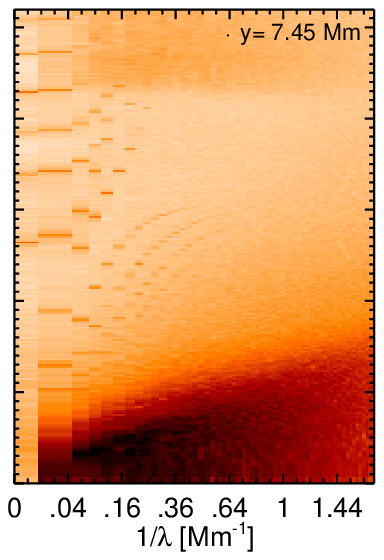}
\includegraphics[width=3.8cm]{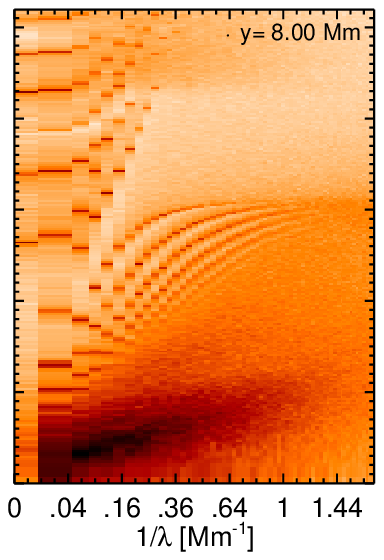}
\includegraphics[width=3.8cm]{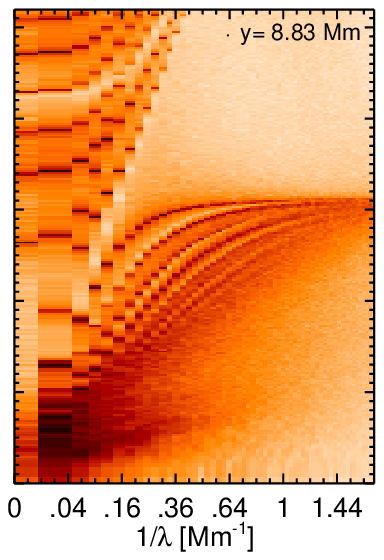}
\figcaption{\label{fig:komega_lc0gg_seq} $k$-$\omega$ diagrams for
various heights of run lc0gg.  The sample heights lie in the middle of
the lower stable region, just below the entropy jump, inside and above
the jump (top row from left to right); in the middle of the convection
zone, near the top of convection zone, at the bottom of the upper
stable zone, close to the middle of this zone (bottom row from left to
right).}
\end{figure*}

The inconspicuous convective signature is the inclined flattened
streak in the very lower left corner at $\lambda$$>$1\,\mem{Mm},
$t$$>$100\,\mem{s}.  To exhibit the location of the convective signal
in the $k$-$\omega$ diagrams more clearly
Fig.~\ref{fig:komega_lc0gg_seq} shows only the lower left corner of
the entire diagram, with a non-linear scale that magnifies the region
corresponding to long time-scales and large wavelength even further.
The figure shows a sequence of $k$-$\omega$ diagrams for various
heights inside the stable regions and the convection zone.

All plots show p-modes with varying relative power.
Inside the convection zone ($y$=4.70\,\mem{Mm})
they are barely visible above the huge background of the convective motions.
However, in the stable layers close to the top of the computational domain
they have almost the same total power as the gravity waves.
The top panels in Fig.~\ref{fig:komega_lc0gg_seq} show a single ridge
between p-modes and g-modes: the fundamental mode (f-mode).

The gravity waves in the lower cavity (four top panels) have a higher
Brunt-V{\"a}is{\"a}l{\"a}-frequency
than the modes in the upper cavity.
Both wave families can be traced into the convection zone
($y$=1.79\,\mem{Mm} and $y$=7.45\,\mem{Mm}).

The extended convective blob is visible in all panels,
reaching from the convection zone, where it dominates the power,
far into the stable regions.
A $k$-$\omega$ diagram even closer to the top than the last
panel in Fig.~\ref{fig:komega_lc0gg_seq} does not show a clear signature
of the convective velocities anymore (the location of the last panel
was chosen accordingly).

While the convective flow overshoots into the stable regions,
the stable regions ``shoot back'', they tunnel somewhat into
the region with the essentially flat entropy profile.  This overlap of
flows leads to the mixing of high-entropy material into the convection
zone, as described in Sect.~\ref{sec:evolconv}.  The region where
convective flows and gravity waves can interact is vertically extended
and not confined just to the top layer of the convection zone.  During
the onset phase of convection, gravity waves are excited in the upper
stable region well before the convective upflows reach the upper
boundary of the entropy plateau.  Convective elements (plumes) do not
just ``smash'' against the boundary.  Instead, the braking of the flow
(and the excitation of the waves) occurs over some extended region.
Rising co-moving fluid packets (adiabatically) compress the material
between themselves and the top boundary.  The pressure rises, brakes
the vertical movement of the convective element, and accelerates
material horizontally.  As a result, the vertical motion of the
convective element is turned into a horizontal flow and the
over-pressure at the top lifts the boundary of the entropy plateau
slightly.  The material in the stable region is pushed away both
horizontally and vertically.  The reaction of the stable layers has an
oscillatory component -- the g-modes.  However, as long as the local
over-pressure from the convection zone below exists, some distortion
remains in the stable layers. The distortion closely follows the
time-evolution of the convective flow and has the same signature in a
$k$-$\omega$ diagram.

While the picture as described above remains qualitatively the same
for all resolutions and heating rates in our grid, there are some
differences and trends visible.  The position of a mode peak in a
$k$-$\omega$ diagram remains essentially the same for all runs since
it only depends on the background stratification which only changes
slightly towards the end of the lc0cg run (largest heating rate).
However, more heating means higher convective velocities and smaller
time scales.  Therefore, the relation between time-scales of
convection and modes changes.  The convective blob moves closer to
the ridges of the g-modes, and a separation becomes more difficult.

The mode peaks appear sharper at lower heating rates, where less
interaction with convection and less change in the background
stratification lead to longer mode lifetimes and ``cleaner'' peaks.
Convective scales are slightly smaller at lower heating rates,
indicated by a shifted maximum and stronger tail towards higher
frequencies.

An increase in resolution does not change the frequencies of already
visible modes noticeably.  However, more and more ridges appear and
this indicates that the wide and flat g-mode nodes are not
well enough resolved.

\subsection{Mixing}
\label{sec:mixing}

Our mixing analysis is preliminary. Mixing inside the convective
region is directly correlated with the averaged vertical velocities
(top panels \abb{fig:stat}). An important feature of these velocity
profiles is the substantial decline of the velocities within the
convective region, starting at a position several hundred $\mem{km}$
above the bottom and below the top of the convection layer. These
plots indicate that the decay of the vertical velocity field is
exponential. Outside the convective region, the velocity amplitudes of
the g-modes dominate the averaged vertical velocity profile. Due to
the wave nature of g-mode oscillations, these rather large velocities
correspond to mixing that is orders of magnitude less efficient than
that inside the convection zone, and possibly negligible far away from
the convection zone. The fact that the g-mode velocity amplitudes
superimpose onto the decay of the convective velocity field
complicates deriving the effective mixing in and into the stable
layers. A preliminary analysis of the oscillation modes
(\kap{sec:modes}) suggests that the convective velocity field and therefor the
corresponding mixing continues the exponential decay into the stable
layer. However, future analysis has to reveal the extent of the
overshooting plumes interact with the horizontal motions of the
g-modes, potentially leading to increased mixing in the stable layers
beyond the convective boundary.

We have performed two test runs with enhanced heating rate and low to
moderate resolution (lc2df and lc2dg, \abb{fig:grid}) and with three
individual fluids in the three different layers of the setup.  The
fluids are treated separately by the code but have the same
properties, i.e.\ the same molecular weight. We use these fluids as
passive tracer particles to establish an approximate upper limit on
mixing \citep{hurlburt:94}. We find that there is a finite mixing of
material across both boundaries. In both runs at both boundaries there
are typically four horizontal layers involved in the main abundance
transition. Mixing from the top stable layer into the convection zone
seems to be more efficient for the higher resolution run. The top
layer abundance reaches a mass fraction of approximately $10^{-5}$
inside the convection zone, whereas the lower resolved simulation
(lc2df) reaches a few $10^{-6}$. Mixing of material into the
convection zone from below is at a similar level but shows a
significant upward trend at late times (after several convective
turn-over times). Mixing of material from below the convection zone is
roughly at the same level as mixing from above, but more efficient for
the lower resolution case than for the higher resolution case.

Mixing at the top unstable-stable interface can be seen in the model
lc0dh with large heating rate and high resolution (cf.\ the bottom
panel in Fig.~\ref{fig:lc0dX_sfluct_reslate}).  The shear flows at the
top of the convection zone ``peel off'' high-entropy material from the
stable layers above.  This material keeps its identity for some time,
remains hotter than the surroundings, and is visible as bright spots
in the center of some vortices. The rather long survival time of these
vortices may be a 2D artifact.

The material in the lower stable zone is too dense
to be mixed directly into the convection zone.
However, the entropy (and density) jump is smoothed out over time.
Then the flow at the bottom of some plumes becomes strong
enough to lift up slightly cooler material from below
that is finally mixed into the convection zone.
It is visible as the dark areas directly beneath the plumes
in Fig.~\ref{fig:lc0dX_sfluct_reslate}.

To some degree models with large heating rate and low resolution seem
to have similar properties as models with smaller heating rate and
higher resolution.  We will therefore continue to perform
simulations with too large heating rates in addition to the runs with
more realistic parameters.

\section{Discussion and conclusions}
\label{sec:concl}

We presented the first set of hydrodynamic simulations specifically
addressing the conditions in the convection zone close the peak
luminosity of the He-shell flash. Compared to previous calculations of
other stellar convection zones our simulations reveal some important
differences. One of the most important is the stiffness of the
convective boundaries of He-shell flash convection in contrast to the
shallow surface convection of A-stars and white dwarfs as studied by
\citet{freytag:96}. Other examples include solar-type convection
\citep{dintrans:05} or much of the parameter range in relative
stability between the unstable and stable layers studied by
\citet{rogers:05}.  Even in the simulations of core convection in
A-type stars \citep{browning:04} convective plumes can cross the
convective boundary significantly. In shallow surface convection zones
the convective downdrafts easily cross the convection boundary
and penetrate deep into the stable layers beneath. No significant
g-mode excitation has been detected by \citet{freytag:96}, which is
probably due to the absence of a stiff resonance floor with which
convective plumes and the pressure field interact. However, it seems
that because of this particular feature of the shallow surface
convection zone \citet{freytag:96} were able to study the decay of
the convective velocity field in isolation. In our He-shell flash
convection we do see exponential decay of the convective velocity
field as well. However it already starts inside the convection zone,
and at or beyond the border quickly intermingles with the effects of
the gravity waves.

The comparison to the new O-shell convection simulations by \citet[][and 2006b, in prep]{maekin:06a}
confirms the importance of g-modes for mixing and to correctly account
for the physics at the convective boundary. Similar to our simulations
their convective boundaries are very stiff, with overshooting of
convective plumes into the stable layer reduced to a minimum. More
important are horizontal wave motions which eventually induce
turbulent mixing. A partial mixing zone at the convective boundaries
is minimal. The O-shell convection simulations include
$\mu$-gradients, contrary to our similations. $\mu$ gradients increase
the impenetrable character of the convective boundaries significantly,
and we plan to include this effect in our simulations in the future.

We tentatively compare the mixing length convective velocity profile
from our 1D stellar evolution template model with the averaged
vertical velocities in the hydrodynamic simulation with realistic
heating rate in \abb{fig:v_hdmlt}. Obviously, there are many
differences in the assumptions of the two calculations, including
geometry, micro-physics and of course the treatment of convection
itself. Nevertheless, we find that the absolute value of the
velocities inside the convection zone are similar. We do
observe a different profile in the two representations. The
hydrodynamic simulation show a more pronounced slope in the velocity
profile within the convection zone than the MLT profile. The peak in
the velocity profile from the hydrodynamic simulation is close to the
bottom of the convection zone. The most important difference is the
significant decline of the velocity field already inside the
convection zone. Near the lower boundary the vertical velocity is
almost 3 orders of magnitude lower than the peak value. This may have
implications for s-process nucleosynthesis as nuclei are on average
exposed much longer to the hottest temperature at the very bottom of
the convection zone. This should be considered in particular for cases
that involve branchings sensitive to the convective time-scale of
He-shell flash convection, like \xeac\ \citep{reifarth:04}.

Consistent with previous work on stellar interior convection we
observe a rich spectrum of g-modes excited in the stable layers. These
g-modes interact with the decaying convective flows at the boundary of
the unstable region. Our simulations show that He-shell flash
convection displays the main ingredients observed in previous stellar
convection simulations, but with different relative contributions of
the various components. The actual overshooting of convective plumes
across the convective boundaries is much smaller than in shallow
surface convection and O-shell convection, while the excitation of
internal gravity is probably stronger because the boundaries are
stiffer.  Although there is a significant overlap of the two processes
we have shown that the analysis of the oscillation modes by means of
the $k$-$\omega$ diagram allows us to disentangle the different
contributions to the velocities. With this approach we observe that
there is a finite mixing of convective motions accross the convective
boundary. We postpone quantifying this mixing to future work. In
addition, we plan to quantify the contribution from wave mixing at
He-shell flash convection boundaries.

We have identified a number of numerical problems at low heating rates
and low resolution that need to be dealt with in future 3D runs. These
include the diffusion of entropy from above into the top zones of the
unstable layer, impeding the development of the correct hydrodynamic
flow-pattern. Our convergence and heating rate study shows that
simulations with larger heating rate are in some ways equivalent to
runs with higher resolution. Properties like the exponential decay
rate of the convective velocity field seem to be only weakly dependent
on the heating rate. However, the excitation of g-modes and their
amplitudes depends sensitively on both the resolution and the heating
rate. While we do see signs of convergence for the convective properties of our
simulations, this is not the case for the g-modes.

We consider 2D vs.\ 3D systematics, the effect of $\mu$-gradients, a
realistic treatment of nuclear energy generation, and a more detailed
study of the interaction of wave mixing with convective overshooting
as our most immediate tasks for future improvements. In addition we
can make our simulations more realistic by adopting an apppropriate
stellar equations of state, by accounting for the spherical geometry
of the He-shell, and by including a realistic gravitational potential.

\acknowledgements We want to thank Kunegunda Belle, Chris Freyer,
Alexander Heger, Robert Stein and Paul Woodward who have supported
this work in various and very generous ways.  This work is in part a
result of Los Alamos National Laboratory's participation in the Joint
Institute for Nuclear Astrophysics (JINA), an NSF Physics Frontier
Center.  This work was funded under the auspices of the U.S.\ Dept.\
of Energy under the ASC program at Los Alamos National Laboratory.


\begin{thebibliography}{58}
\expandafter\ifx\csname natexlab\endcsname\relax\def\natexlab#1{#1}\fi

\bibitem[{Alexander \& Ferguson(1994)}]{alexander:94}
Alexander, D. \& Ferguson, J. 1994, ApJ, 437, 879

\bibitem[{{Althaus} {et~al.}(2005){Althaus}, {Serenelli}, {Panei},
  {C{\'o}rsico}, {Garc{\'{\i}}a-Berro}, \& {Sc{\'o}ccola}}]{althaus:05}
{Althaus}, L.~G., {Serenelli}, A.~M., {Panei}, J.~A., {C{\'o}rsico}, A.~H.,
  {Garc{\'{\i}}a-Berro}, E., \& {Sc{\'o}ccola}, C.~G. 2005, A\&A, 435, 631

\bibitem[{{Angulo} {et~al.}(1999){Angulo}, {Arnould}, \& {Rayet, M. et
  al.}}]{angulo:99}
{Angulo}, C., {Arnould}, M., \& {Rayet, M. et al.} 1999, Nucl.\ Phys., A 656,
  3, {NACRE} compilation

\bibitem[{{Arlandini} {et~al.}(1999){Arlandini}, {K\"appeler}, {Wisshak},
  {Gallino}, {Lugaro}, {Busso}, \& {Straniero}}]{arlandini:99}
{Arlandini}, C., {K\"appeler}, F., {Wisshak}, K., {Gallino}, R., {Lugaro}, M.,
  {Busso}, M., \& {Straniero}, O. 1999, ApJ, 525, 886

\bibitem[{{Asida} \& {Arnett}(2000)}]{asida:00}
{Asida}, S.~M. \& {Arnett}, D. 2000, ApJ, 545, 435

\bibitem[{Baltrusaitis {et~al.}(1996)Baltrusaitis, Gittings, Weaver, Benjamin,
  \& Budzinski}]{baltrusaitis:96}
Baltrusaitis, R., Gittings, M., Weaver, R., Benjamin, R., \& Budzinski, J.
  1996, Phys. Fluids, 8, 2471

\bibitem[{{Bazan} \& {Arnett}(1998)}]{bazan:98}
{Bazan}, G. \& {Arnett}, D. 1998, ApJ, 496, 316

\bibitem[{{Beers} \& Christlieb(2005)}]{beers:05}
{Beers}, T.~C. \& Christlieb, N. 2005, ARAA, 43, xxx

\bibitem[{Belle {et~al.}(2005)Belle, Coker, \& Hueckstaedt}]{belle:05}
Belle, K., Coker, R., \& Hueckstaedt, R. 2005, ApJ

\bibitem[{Bl\"ocker(1995)}]{bloecker:95a}
Bl\"ocker, T. 1995, A\&A, 297, 727

\bibitem[{B\"ohm-Vitense(1958)}]{boehm-vitense:58}
B\"ohm-Vitense, E. 1958, Z.\ Astrophys., 46, 108

\bibitem[{{Browning} {et~al.}(2004){Browning}, {Brun}, \&
  {Toomre}}]{browning:04}
{Browning}, M.~K., {Brun}, A.~S., \& {Toomre}, J. 2004, ApJ, 601, 512

\bibitem[{{Brun}(2004)}]{brun:04b}
{Brun}, A.~S. 2004, Solar Phys., 220, 333

\bibitem[{{Busso} {et~al.}(1999){Busso}, {Gallino}, \& {Wasserburg}}]{busso:99}
{Busso}, M., {Gallino}, R., \& {Wasserburg}, G.~J. 1999, ARA\&A, 37, 239

\bibitem[{{Cristallo} {et~al.}(2004){Cristallo}, {Gallino}, \&
  {Straniero}}]{cristallo:04}
{Cristallo}, S., {Gallino}, R., \& {Straniero}, O. 2004, Memorie della Societa
  Astronomica Italiana, 75, 174

\bibitem[{Dendy \& Gittings(2005)}]{dendy:05}
Dendy, E. \& Gittings, M. 2005, Los Alamos report LA-UR-05-1642

\bibitem[{Denissenkov \& Herwig(2003)}]{denissenkov:03a}
Denissenkov, P.~A. \& Herwig, F. 2003, ApJ Lett., 590, L99

\bibitem[{{Dintrans} {et~al.}(2005){Dintrans}, {Brandenburg}, {Nordlund}, \&
  {Stein}}]{dintrans:05}
{Dintrans}, B., {Brandenburg}, A., {Nordlund}, {\AA}., \& {Stein}, R.~F. 2005,
  A\&A, 438, 365

\bibitem[{Freytag {et~al.}(1996)Freytag, Ludwig, \& Steffen}]{freytag:96}
Freytag, B., Ludwig, H.-G., \& Steffen, M. 1996, A\&A, 313, 497

\bibitem[{{Freytag} {et~al.}(2002){Freytag}, {Steffen}, \&
  {Dorch}}]{freytag:02}
{Freytag}, B., {Steffen}, M., \& {Dorch}, B. 2002, Astr.\ Nachr., 323, 213

\bibitem[{{Geisler} {et~al.}(2005){Geisler}, {Smith}, {Wallerstein},
  {Gonzalez}, \& {Charbonnel}}]{geisler:05}
{Geisler}, D., {Smith}, V.~V., {Wallerstein}, G., {Gonzalez}, G., \&
  {Charbonnel}, C. 2005, AJ, 129, 1428

\bibitem[{Herwig(2000)}]{herwig:99a}
Herwig, F. 2000, A\&A, 360, 952

\bibitem[{Herwig(2004)}]{herwig:03c}
---. 2004, ApJ, 605, 425

\bibitem[{{Herwig}(2004)}]{herwig:04a}
{Herwig}, F. 2004, ApJS, 155, 651

\bibitem[{{Herwig} \& {Austin}(2004)}]{herwig:04b}
{Herwig}, F. \& {Austin}, S.~M. 2004, ApJ Lett., 613, L73

\bibitem[{Herwig {et~al.}(1997)Herwig, Bl\"ocker, Sch\"onberner, \& {El
  Eid}}]{herwig:97}
Herwig, F., Bl\"ocker, T., Sch\"onberner, D., \& {El Eid}, M.~F. 1997, A\&A,
  324, L81

\bibitem[{Holmes {et~al.}(1999)Holmes, Dimonte, Fryxell, Gittings, Grove,
  Schneider, Sharp, Velikovich, Weaver, \& Zhang}]{holmes:99}
Holmes, R., Dimonte, G., Fryxell, B., Gittings, M., Grove, J., Schneider, M.,
  Sharp, D., Velikovich, A., Weaver, R., \& Zhang, Q. 1999, J. Fluid Mech.,
  389, 55

\bibitem[{Hueckstaedt {et~al.}(2005)Hueckstaedt, Batha, Balkey, Delamater,
  Fincke, Holmes, Lanier, Magelssen, Scott, Taccetti, Horsfield, Parker, \&
  Rothman}]{hueckstaedt:05}
Hueckstaedt, R., Batha, S., Balkey, M., Delamater, N., Fincke, J., Holmes, R.,
  Lanier, N., Magelssen, G., Scott, J., Taccetti, J., Horsfield, C., Parker,
  K., \& Rothman, S. 2005, Ap\&SS, 298, 255

\bibitem[{{Hurlburt} {et~al.}(1984){Hurlburt}, {Toomre}, \&
  {Massaguer}}]{hurlburt:84}
{Hurlburt}, N.~E., {Toomre}, J., \& {Massaguer}, J.~M. 1984, \apj, 282, 557

\bibitem[{Hurlburt {et~al.}(1986)Hurlburt, Toomre, \& Massaguer}]{hurlburt:86}
Hurlburt, N.~E., Toomre, J., \& Massaguer, J.~M. 1986, ApJ, 311, 563

\bibitem[{Hurlburt {et~al.}(1994)Hurlburt, Toomre, Massaguer, \&
  Zahn}]{hurlburt:94}
Hurlburt, N.~E., Toomre, J., Massaguer, J.~M., \& Zahn, J.-P. 1994, ApJ, 421,
  245

\bibitem[{Iben \& Renzini(1983)}]{iben:83b}
Iben, Jr., I. \& Renzini, A. 1983, ARA\&A, 21, 271

\bibitem[{Iglesias \& Rogers(1996)}]{iglesias:96}
Iglesias, C.~A. \& Rogers, F.~J. 1996, ApJ, 464, 943

\bibitem[{Langer {et~al.}(1985)Langer, {El Eid}, \& Fricke}]{langer:85}
Langer, N., {El Eid}, M., \& Fricke, K.~J. 1985, A\&A, 145, 179

\bibitem[{{Lugaro} {et~al.}(2003{\natexlab{a}}){Lugaro}, {Davis}, {Gallino},
  {Pellin}, {Straniero}, \& {K{\" a}ppeler}}]{lugaro:02b}
{Lugaro}, M., {Davis}, A.~M., {Gallino}, R., {Pellin}, M.~J., {Straniero}, O.,
  \& {K{\" a}ppeler}, F. 2003{\natexlab{a}}, ApJ, 593, 486

\bibitem[{{Lugaro} {et~al.}(2003{\natexlab{b}}){Lugaro}, {Herwig}, {Lattanzio},
  {Gallino}, \& {Straniero}}]{lugaro:02a}
{Lugaro}, M., {Herwig}, F., {Lattanzio}, J.~C., {Gallino}, R., \& {Straniero},
  O. 2003{\natexlab{b}}, ApJ, 586, 1305

\bibitem[{Maekin \& Arnett(2006)}]{maekin:06a}
Maekin, C. \& Arnett, D. 2006, ApJ Lett., in press

\bibitem[{Mazzitelli {et~al.}(1999)Mazzitelli, D'Antona, \&
  Ventura}]{mazzitelli:99}
Mazzitelli, I., D'Antona, F., \& Ventura, P. 1999, A\&A, 348, 846

\bibitem[{Merryfield(1995)}]{merryfield:95}
Merryfield, W.~J. 1995, ApJ, 444, 318

\bibitem[{{Nordlund}(1982)}]{nordlund:82}
{Nordlund}, A. 1982, \aap, 107, 1

\bibitem[{{Nordlund} {et~al.}(1997){Nordlund}, {Spruit}, {Ludwig}, \&
  {Trampedach}}]{nordlund:97}
{Nordlund}, A., {Spruit}, H.~C., {Ludwig}, H.-G., \& {Trampedach}, R. 1997,
  A\&A, 328, 229

\bibitem[{{Porter} \& {Woodward}(1994)}]{porter:94}
{Porter}, D.~H. \& {Woodward}, P.~R. 1994, ApJS, 93, 309

\bibitem[{{Porter} \& {Woodward}(2000)}]{porter:00}
---. 2000, ApJS, 127, 159

\bibitem[{Reifarth {et~al.}(2005)Reifarth, Alpizar-Vicente, Hatarik, Bredeweg,
  Esch, Greife, Haight, Kronenberg, O'Donnell, Rundberg, Schwantes, Ullmann,
  Vieira, \& Wouters}]{reifarth:05}
Reifarth, R., Alpizar-Vicente, A., Hatarik, R., Bredeweg, T.~A., Esch, E.-I.,
  Greife, U., Haight, R.~C., Kronenberg, A., O'Donnell, J.~M., Rundberg, R.~S.,
  Schwantes, J.~M., Ullmann, J.~L., Vieira, D.~J., \& Wouters, J.~M. 2005, in
  ASP Conf. Proc. 769, 1323--1326

\bibitem[{{Reifarth} {et~al.}(2004){Reifarth}, {K{\" a}ppeler}, {Voss},
  {Wisshak}, {Gallino}, {Pignatari}, \& {Straniero}}]{reifarth:04}
{Reifarth}, R., {K{\" a}ppeler}, F., {Voss}, F., {Wisshak}, K., {Gallino}, R.,
  {Pignatari}, M., \& {Straniero}, O. 2004, ApJ, 614, 363

\bibitem[{{Renda} {et~al.}(2004){Renda}, {Fenner}, {Gibson}, {Karakas},
  {Lattanzio}, {Campbell}, {Chieffi}, {Cunha}, \& {Smith}}]{renda:04}
{Renda}, A., {Fenner}, Y., {Gibson}, B.~K., {Karakas}, A.~I., {Lattanzio},
  J.~C., {Campbell}, S., {Chieffi}, A., {Cunha}, K., \& {Smith}, V.~V. 2004,
  MNRAS, 354, 575

\bibitem[{{Robinson} {et~al.}(2004){Robinson}, {Demarque}, {Li}, {Sofia},
  {Kim}, {Chan}, \& {Guenther}}]{robinson:04}
{Robinson}, F.~J., {Demarque}, P., {Li}, L.~H., {Sofia}, S., {Kim}, Y.-C.,
  {Chan}, K.~L., \& {Guenther}, D.~B. 2004, MNRAS, 347, 1208

\bibitem[{{Rogers} \& {Glatzmaier}(2005)}]{rogers:05}
{Rogers}, T.~M. \& {Glatzmaier}, G.~A. 2005, \apj, 620, 432

\bibitem[{{Siess} {et~al.}(2002){Siess}, {Livio}, \& {Lattanzio}}]{siess:02}
{Siess}, L., {Livio}, M., \& {Lattanzio}, J. 2002, ApJ, 570, 329

\bibitem[{{Stein} \& {Nordlund}(1989)}]{stein:89}
{Stein}, R.~F. \& {Nordlund}, A. 1989, ApJ Lett., 342, L95

\bibitem[{Stein \& Nordlund(1998)}]{stein:98}
Stein, R.~F. \& Nordlund, A. 1998, ApJ, 499, 914

\bibitem[{{Travaglio} {et~al.}(2004){Travaglio}, {Gallino}, {Arnone}, {Cowan},
  {Jordan}, \& {Sneden}}]{travaglio:04}
{Travaglio}, C., {Gallino}, R., {Arnone}, E., {Cowan}, J., {Jordan}, F., \&
  {Sneden}, C. 2004, ApJ, 601, 864

\bibitem[{{Venn} {et~al.}(2004){Venn}, {Irwin}, {Shetrone}, {Tout}, {Hill}, \&
  {Tolstoy}}]{venn:04}
{Venn}, K.~A., {Irwin}, M., {Shetrone}, M.~D., {Tout}, C.~A., {Hill}, V., \&
  {Tolstoy}, E. 2004, AJ, 128, 1177

\bibitem[{{Ventura} {et~al.}(2000){Ventura}, {D'Antona}, \&
  {Mazzitelli}}]{ventura:00}
{Ventura}, P., {D'Antona}, F., \& {Mazzitelli}, I. 2000, A\&A, 363, 605

\bibitem[{{Ventura} {et~al.}(2002){Ventura}, {D'Antona}, \&
  {Mazzitelli}}]{ventura:02}
---. 2002, A\&A, 393, 215

\bibitem[{{Young} \& {Arnett}(2005)}]{young:05a}
{Young}, P.~A. \& {Arnett}, D. 2005, ApJ, 618, 908

\bibitem[{{Young} {et~al.}(2005){Young}, {Meakin}, {Arnett}, \&
  {Fryer}}]{young:05b}
{Young}, P.~A., {Meakin}, C., {Arnett}, D., \& {Fryer}, C.~L. 2005, ApJ Lett.,
  629, L101

\bibitem[{{Zinner}(1998)}]{zinner:98}
{Zinner}, E. 1998, Ann. Rev. Earth Planet. Sci., 26, 147

\end{thebibliography}

\clearpage

\begin{deluxetable}{lllll} 
\tablecolumns{5} 
\tablewidth{0pc} 
\tablecaption{\label{tab:prop70238}
Properties of  He-shell region in stellar model 70238} 
\tablehead{ 
 & \colhead{box bot.}   & \colhead{conv. bot.}    & \colhead{conv. top} & 
\colhead{box top}  }
\startdata 
$\mem{ D/ (cgs)   }$&     0.000E+00 &     1.634E+12 &    1.051E+12 &   0.000E+00 \\
$\mem{ R/ R_\odot }$&     1.088E-02 &     1.324E-02 &    2.206E-02 &   2.681E-02 \\
$\mem{H_p/ R_\odot}$&     1.296E-03 &     2.439E-03 &    1.258E-03 &   2.337E-03\\
$\mem{M_r/ M_\odot}$&     0.5230    &     0.5704    &    0.5918    &   0.5929 \\
$\mem{\rho/ (cgs) }$&     1.267E+05 &     1.173E+04 &    7.004E+02 &   2.782E+01\\
$\mem{T / K       }$&     1.544E+08 &     2.479E+08 &    4.631E+07 &   4.872E+07\\
$\mem{\ln P (cgs) }$&     48.68     &     46.62     &    42.16     &   39.16  \\  
\enddata 
\end{deluxetable}

\clearpage

\begin{figure}
\figcaption{\label{fig:t-r} Time evolution of the radial location of
the He-shell flash convection zone based on the 1D stellar evolution
model. Time is set to zero at the peak of the He-burning
luminosity. The ordinate is zero at the stellar center. The grey
shaded area represents the He-shell flash convection zone. Dots along
the boundary indicate individual time steps in the 1D evolution model
sequence.  The vertical solid line at $t=-0.07\jahre$ indicates the
position and extent of the hydrodynamic simulation box. Dashed and
dotted lines correpsond to the Lagrangian coordinates given in the
legend, and visualize the expansion of the He-shell as a result of
the He-shell flash. }
\end{figure}

\begin{figure}
\figcaption{\label{fig:t-lhe-p} Time evolution of the He-burning
luminosity and the pressure at the boundary of the convection zone for
the He-shell flash. The time axis, grey shades, and dots have the same
meaning as in \abb{fig:t-r}. }
\end{figure}

\begin{figure}
\plotone{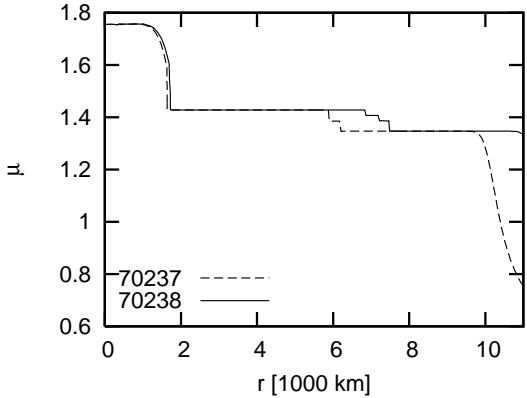} 
\figcaption{\label{fig:r-mu} Mean molecular weight of the fully
ionized material in the simulated shell of the two stellar evolution
models before the peak He-burning flash luminosity. The radius is set
to zero at the stellar radius of $7500\mem{km}$, coinciding with the
bottom of the box for the hydrodynamic simulations. The two steps in
the mean molecular weight are associated with the bottom of the
He-shell flash convection zone at $1650 \mem{km}$ and the top at $7732
\mem{km}$ in model 70238. For model 70237 (dashed line) the bottom of
the H-rich envelope is located at $r=10000\mem{km}$.  }
\end{figure}

\begin{figure}
\figcaption{\label{fig:grid}The grid in resolution and heating rate of
  the standard 2D simulation set lc0. In the figure captions the
  combination of two letters identifies the simulation in this grid.
  The first letter gives the heating rate, the second the resolution.
  For example, run df has a heating rate of 30 times the standard case
  and a 300x100 grid. The numbers in the table give the length of the
  run in seconds. The standard heating rate g corresponds to what is
  described in \kap{kap:multi-D-setup}.}
\end{figure}


\clearpage
\begin{figure*}

  \figcaption{\label{fig:lc0gh_prelfluct_single} Fully developed
    convection in a high-resolution 2D run with standard heating rate.
    The flow field is represented by 25x75 pseudo-streamlines,
    integrated over 100\,\mem{s} for the constant velocity field of
    this single snapshot of evolved convection during the lc0gh run.
    The color indicates the corresponding pressure inhomogeneities.
    The pressure itself does not deviate much from the initial
    values.  Therefore, to make the fluctuations visible, the
    \emph{horizontal average} of the pressure has been subtracted from
    the pressure value of every grid point.  Bright means
    over-pressure (prominently where a ``mushroom'' approaches the
    upper boundary of the unstable region: compare with
    Fig.~\ref{fig:lc0gh_sfluct_single}).  Dark color indicates
    low-pressure (in the ``eyes'' of the vortices).  The boundaries of
    the unstable layer (the entropy plateau in
    Fig.~\ref{fig:qinit_xc3}) at $y$=1.7\,\mem{Mm} and
    $y$=7.7\,\mem{Mm} are clearly marked in the flow field and the
    pressure inhomogeneities.  }
\end{figure*}

\begin{figure*}

  \figcaption{\label{fig:lc0gh_sfluct_single} Entropy inhomogeneities
    for the same model of the lc0gh sequence as in
    \abb{fig:lc0gh_prelfluct_single}.  Again, the horizontal
    average of the entropy has been subtracted to render the small
    fluctuations visible.  A bright color indicates material with
    higher entropy (and in fact higher temperature -- due to the near
    pressure-equilibrium).  Dark means low entropy (or temperature).
    The boundaries of the entropy plateau at $y$=1.7\,\mem{Mm} and
    $y$=7.7\,\mem{Mm} are again clearly visible in the change of the
    patterns.  The subtraction of the horizontal mean causes bright
    features to be accompanied by dark horizontal stripes.  These are
    pure artifacts of the visualization procedure and do not exist in
    the simulation data itself.  }
\end{figure*}

\clearpage

\begin{figure}
\figcaption{\label{fig:lc0gh_sfluct_onset} Entropy inhomogeneities during the onset of convection for the lc0gh run.}
\end{figure}

\begin{figure}
\figcaption{\label{fig:multiplotfield_lc0gh_onset_P-relfluct_vstream} Like \abb{fig:lc0gh_prelfluct_single} for an earlier time during the onset of convection. Note the build-up of higher pressure regions above rising plumes that can be associated with the excitation of gravity waves in the top stable layer. }
\end{figure}

\begin{figure}
  \figcaption{\label{fig:t-vrms} Time evolution of the
  maximum vertical, rms-averaged velocity for the standard heating
  rate g (top panel) and the enhanced heating rage d (bottom panel)
  for a range of resolutions.}
\end{figure}
\begin{figure}
  \figcaption{\label{fig:lc0dX_sfluct_researly} Resolution sequence of
    entropy inhomogeneities during the onset of convection.}
\end{figure}

\begin{figure}
  \figcaption{\label{fig:lc0Xg_sfluct_heatearly} Heating rate sequence
    (decreasing from top to bottom) of entropy inhomogeneities during
    the onset of convection.  The time steps are scaled with $q_{\rm
      c}^{-0.21}$.  The color palettes are scaled with $q_{\rm
      c}^{0.5}$.}
\end{figure}

\clearpage

\begin{figure}
\figcaption{\label{fig:lc0gh_sfluct_evol} Time-sequence of 
the entropy fluctuations for evolved convection, lc0gh run.}
\end{figure}

\clearpage

\clearpage

\begin{figure}
\figcaption{\label{fig:lc0dX_sfluct_reslate} Resolution sequence and large heating rate,  entropy inhomogeneities for evolved convection.}
\end{figure}

\begin{figure}
  \figcaption{\label{fig:lc0gX_sfluct_reslate} Resolution sequence and
    realistic heating rate, entropy inhomogeneities for evolved
    convection.}
\end{figure}

\begin{figure}
  \figcaption{\label{fig:lc0Xg_sfluct_heatlate} Heating rate sequence
    of the entropy inhomogeneities for evolved convection. The colors
    are scaled with $q_{\rm c}^{0.5}$.}
\end{figure}

\begin{figure}
\figcaption{\label{fig:lc0Xg_pfluct_heatlate} Pressure  inhomogeneities and pseudo-streamlines for evolved convection
for runs with different heating rates.
The colors are scaled with $q_{\rm c}^{0.5}$,
the integration time for the pseudo-streamlines with $q_{\rm c}^{0.5}$.}
\end{figure}

\clearpage

\clearpage

\begin{figure}
\includegraphics[width=7.6cm]{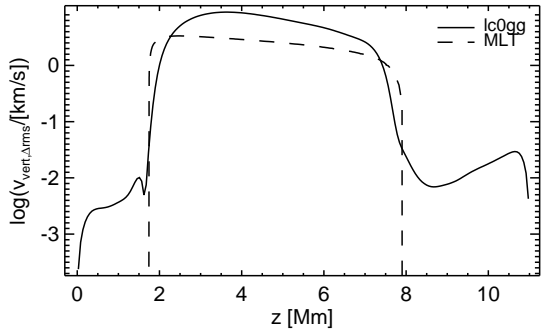}
\figcaption{\label{fig:v_hdmlt}
Comparison of vertical velocities derived from a hydrodynamics model
(continuous lines: lc0gg)
and mixing-length velocities from stellar evolution calculations
(dashed lined: model 70238).}
\end{figure}

\end{document}